# Distinguishing Thixotropy from Viscoelasticity

Mayank Agarwal,[a] Shweta Sharma,[a] V. Shankar*, Yogesh M. Joshi*

Department of Chemical Engineering,

Indian Institute of Technology Kanpur, Kanpur 208016. INDIA.

* Corresponding authors, email id: vshankar@iitk.ac.in and joshi@iitk.ac.in

[a] Both authors contributed equally to this manuscript.

**Abstract:**

Owing to non-linear viscoelasticity, materials often show characteristic features that resemble those of thixotropy. This issue has been debated in the literature over the past several decades, and several experimental protocols have been proposed to distinguish thixotropy from viscoelasticity. In this work, we assess these protocols by carrying out experiments using polymer solutions, thixotropic clay dispersions and modelling their behavior respectively using the FENE-P constitutive equation and a viscoelastic aging model. We find that the criteria proposed in the literature, such as a step-down jump in shear rate and shear start-up at different waiting times elapsed since pre-shear, are inadequate to distinguish thixotropy from viscoelasticity. In marked contrast, we show that the application of step-strain or step-stress after cessation of pre-shear serves as a useful discriminant between thixotropy and viscoelasticity. In thixotropic materials, we observe that the application of step strain (or step stress) after cessation of pre-shear eventually leads to slowing down of relaxation dynamics as a function of waiting time. However, for viscoelastic materials, the relaxation modulus (creep compliance) curve shifts to lower modulus (higher compliance) values as a function of waiting time until equilibrium is reached. While the proposed criterion offers a robust distinction between viscoelasticity and thixotropy for the systems studied here, further experimental investigations based on other systems are needed to establish its versatility and will lead to a greater insight into this long-standing issue in rheological categorization.

# I. Introduction

Viscoelasticity stems from the ability of a material to store energy over an observation timescale. On the other hand, thixotropy concerns time dependent evolution of the material's viscosity and how the deformation field affects the same. Viscoelastic materials are ubiquitous in nature and industry. Such materials include, but are not limited to, polymeric solutions and melts, colloidal dispersions, emulsions, surfactant solutions, a variety of multi-component and multi-phase commercial products, foodstuff, etc. Some of these materials attain an equilibrium state over practical timescales, while some get kinetically arrested in states that are not at equilibrium. Consequently, the former shows time-invariant rheological properties once equilibrium is established, while the latter shows slow but continual evolution of the same, thereby demonstrating time-dependent behavior. Irrespective of whether a material reaches equilibrium or not, upon application or removal of the deformation field, any viscoelastic material exhibits a transient whose timescale depends on its relaxation time. Often, rheological behavior during a transient is reminiscent of thixotropic behavior of a material [1-3]. With an increase in relaxation time, it becomes exceedingly difficult to distinguish between intrinsic viscoelasticity and thixotropy. To this end, various protocols have been proposed in the literature to differentiate between viscoelasticity and thixotropy. In this work, we critically evaluate these protocols using representative experimental systems as well as theoretical models and analyze the current state of understanding in this subject.

The term "Thixotropy" was introduced by Freundlich [4] in 1938, but the origin of the concept was given by Peterefi [5] in 1927. It is derived from the Greek words; θίξις (thixis: stirring, shaking) and τόπος (trepo: turning or changing) and referred to as gel-sol transition induced by mechanical agitation. Thixotropy has been an active area of research, particularly in terms of its definition, measurement techniques and theoretical framework. The most accepted definition of thixotropy is given by IUPAC [6]: *"The application of a finite shear to a system after a long rest may result in a decrease of the viscosity or the consistency. If the decrease persists when the shear is discontinued, this behaviour is called work softening (or shear breakdown), whereas if the original viscosity or consistency is recovered this behaviour is called thixotropy."* After discussing various definitions of thixotropy, Barnes [1] in his 1997 review paper mentions, *"All liquids with microstructure can show thixotropy, because thixotropy only reflects the finite time taken to move* *from any one state of microstructure to another and back again, whether from different states of flow or to or from rest."* Moreover, materials which exhibit shear thinning demonstrate a decrease in viscosity with time in a shear start-up flow associated with corresponding shear rates. On the other hand, the viscosity increases gradually upon decrease in deformation rate or upon cessation of flow. Considering this, Barnes [1] states that all shear thinning systems necessarily show a thixotropic response as a finite time is always required to produce rearrangements necessary to obtain shear thinning. Very recently, Rubio-Hernández et al. [7] carried out experiments on fumed silica suspension and proposed that a relationship does exist between shear thinning and thixotropy. The relationship between shear thinning behavior and the ambiguous signature of thixotropy it presents, has also been debated in the literature. An early review by Mewis [8] and more recent reviews by Mewis and Wagner [2, 9] noted that a decrease in viscosity due to an increase in shear rate cannot be ascribed to thixotropy, because a non-linear viscoelastic system showing shear thinning behavior can also exhibit the same. Therefore, it is important to have a clear distinction between thixotropy and viscoelasticity (including for viscoelastic materials that show non-linear effects such as shear thinning). This, however, is a difficult task as an unambiguous definition of thixotropy is required that excludes viscoelasticity. To begin with, in the absence of any explicit classification, the simplest approach is to exclude from the definition of the required conditions of thixotropy, all those effects that are exhibited by usual viscoelastic materials such as polymer melts and solutions and corresponding constitutive equations that are universally accepted to be non-thixotropic. We shall revisit this issue below after discussing how this aspect is addressed in the literature.

Over past two decades there has been a significant surge in understanding thixotropic and yield stress behavior of soft materials, both theoretically and experimentally [10-23]. In the literature, several experimental procedures have been adopted/proposed to assess the presence of thixotropy in a material. One of the most popular experimental protocols which is used to gauge the extent of thixotropy is the rheological hysteresis experiment. This technique was introduced by Green and Weltman in 1943 [24], and later it is adopted by several authors [11, 15, 25]. In this technique, the shear rate is first increased (decreased) from a very small (large) value and then decreased (increased) with the same rate to the lower (higher) value. The increase and decrease can be a stepwise or continuous ramp. When the transient shear stress is plotted against the shear rate, a thixotropic material typically shows a hysteresis loop. The area of this loop, in principle, gives a measure of the extent of thixotropy but depends on the values of maximum and minimum shear rate and the rate of change of shear rate. However, a disadvantage of this test is the coupling between the shear rate and time effects. Interestingly, the presence of a hysteresis loop is also reported for many viscoelastic materials via experiments as well as theoretical calculations [26-28]. The very fact that a viscoelastic system will always show hysteresis when the timescale probed is below its relaxation time, and since the relaxation time of a viscoelastic material can vary from nanoseconds to centuries [29], the hysteresis loop test cannot definitively differentiate thixotropy and viscoelasticity.

In order to examine/ascertain the effect of viscoelasticity on a thixotropic fluid, Dullaert and Mewis [30] studied the transient stress response in a fumed silica suspension after subjecting it to shear rate step-down jump from a high value. The resulting stress transient showed an initial decrease in stress, which corresponds to the viscoelastic relaxation, and this was followed by an increase in stress. The authors attributed the rise in stress to microstructural build-up. Historically, the step-down in shear rate experiment was introduced by Weltman [24] to study the thixotropy in pigment suspensions. Mewis and Wagner [2] proposed this method, which is schematically shown in Fig. 1, as a tool to distinguish among various rheological responses. In this protocol, as shown in Fig. 1(a), a material is subjected to a shear rate step-down jump from the initial shear rate ($\dot{\gamma}_i$) to the final shear rate ($\dot{\gamma}_e$), and the subsequent stress response is monitored as a function of time. They mention that the suggested protocol is advantageous over rheological hysteresis as the shear rate and time can be varied independently. They propose that for the usual viscoelastic fluids, irrespective of whether it is in the linear or non-linear region, resultant stress decreases monotonically and reaches a new plateau, as shown in Fig. 1(b). The stress for ideal (inelastic) thixotropic fluids, on the other hand, drops instantaneously to a lower value and then increases gradually to achieve a new steady-state, as depicted in Fig. 1(c). However, for viscoelastic-thixotropic materials, they proposed that the response is a combination of viscoelastic and inelastic thixotropic responses, wherein the stress first reduces and then increases with time, as shown in Fig. 1(d). There are many reports available in the literature, which confirm the above protocol for various thixotropic systems such as bentonite-water suspension [31], crude oil [32], fumed silica suspension [30, 33-35], carbon black suspension [36], Carbopol dispersion [37], human blood [38], theoretical models [39, 40], which suggest this distinction to be useful. Recently Choi et al. [41]

discussed the role of elasticity during step down shear rate experiments using 2.9 vol % fumed silica suspension. The authors observed that typical stress build-up observed during step down shear rate experiments for thixotropic materials cannot always be treated as a pure viscous response as the stress build-up is also associated with a notable elastic stress evolution. However, it is necessary to ascertain whether universally accepted non-thixotropic viscoelastic materials such as polymer melts and solutions, and constitutive equations that describe them, do not exhibit behavior similar to so-called `viscoelastic-thixotropic' materials, in order to avoid incorrect assessments.

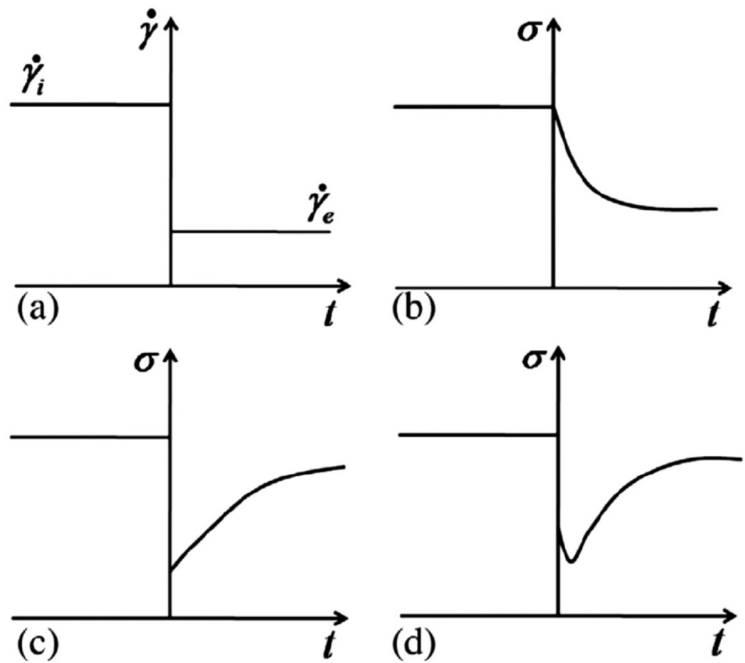

**Fig. 1**. For a flow field wherein the shear rate undergoes a step-down jump as shown in (a), various model responses have been categorized by Mewis and Wagner [9] as, b) viscoelastic; c) inelastic thixotropic; d) viscoelastic thixotropic. (Reprinted from Advances in Colloid and Interface Science, 147, Mewis, J., and Wagner, N. J., Thixotropy, 214-227, Copyright (2009), with permission from Elsevier).

Interestingly, in thermodynamically out of equilibrium soft materials, also known as soft glassy materials (SGMs), the relaxation time increases as a function of time under quiescent conditions, a behavior known as physical aging. On the other hand, the relaxation time of SGMs decreases under the application of the deformation field, a phenomenon termed as rejuvenation.

This behavior of SGMs, corresponding to an increase and decrease in viscosity respectively under quiescent conditions and under application of the deformation field, is reminiscent of thixotropic behavior [2, 9, 18]. Physical aging behavior of SGMs is assessed by subjecting the same to various deformation fields at different waiting times elapsed since shear melting. For example, if a material is subjected to creep flow, then lesser strain gets induced in the same when applied at higher waiting times. Equivalently, if step strain is applied to an aging material, the subsequent relaxation gets slower at higher waiting times [42-45]. In a transient start-up shear flow under constant rate, on the other hand, SGMs show a stress overshoot whose magnitude increases with increase in the waiting time at which step shear rate jump is applied [46]. In principle, these tests can be used as a confirmatory test to assess part of the thixotropy definition associated with increase in viscosity under quiescent conditions after removal of the deformation field. However, one needs to examine whether above-mentioned universally accepted non-thixotropic viscoelastic materials with high relaxation time also show misleading signatures of physical aging.

Considering the ambiguity in the definition of thixotropy in comparison with viscoelasticity, Larson [29] recently proposed an important distinction by excluding elasticity from the definition of thixotropy. He divided the response that can be termed as thixotropic depending upon the relative magnitude of thixotropic timescale in comparison to viscoelastic relaxation time. The thixotropic timescale ($\tau_{Th}$) is defined as the characteristic time associated with the structural buildup under quiescent conditions. According to Larson [29], it is the timescale that is associated with the relaxation of viscosity or other viscoelastic parameters. However, the corresponding structural restoring forces are not strong enough, leading to stresses that are predominantly viscous. On the other hand, the viscoelastic relaxation time ($\tau_{VE}$) is that timescale over which stress relaxes. Larson defined thixotropic response for which $\tau_{Th} > \tau_{VE}$, such that the inelastic limit $\tau_{VE} \to 0$ (and finite $\tau_{Th}$), leads to `ideal thixotropic' response. Larson and coworkers [3, 29] commented that the response to step-down in shear rate given by the qualitative behavior in Fig 1(d) proposed by Mewis and Wagner [9], where stress decay occurs over a timescale significantly smaller than the timescale over which stress buildup occurs, represents a viscoelastic thixotropic response. Larson and Wei [3] further propose their own tentative criterion according to which state of a material whose viscosity decreases more than half under stress but undergoes less than 1 % strain recovery upon removal of stress can be termed as thixotropic. They state that depending on the strength of deformation field and time, the comparative strength of the thixotropic time scale with respect to viscoelastic relaxation time changes, and hence the same material can show a variety of responses such as viscoelastic, thixotropic, viscoelastic aging, etc. [3]. However, the other important issue with Larson and coworker's [3, 29] distinction is that thermodynamically out of equilibrium soft materials, that show all the characteristic features of thixotropy get naturally excluded from being called thixotropic. This is because, their (viscoelastic) relaxation time continuously evolves, and eventually tends to infinity. Consequently, soon after pre-shear, it has been proposed that the viscoelastic timescale of a soft glassy material becomes greater than thixotropic timescale, and according to Larson, loses its distinction as being thixotropic. In principle, Larson's distinction should have solved the dilemma of distinguishing thixotropy and viscoelasticity. However, in our opinion, Larson's definition of thixotropic response is restrictive and many materials (in their usual states) that are otherwise referred to as thixotropic in the literature will lose that status.

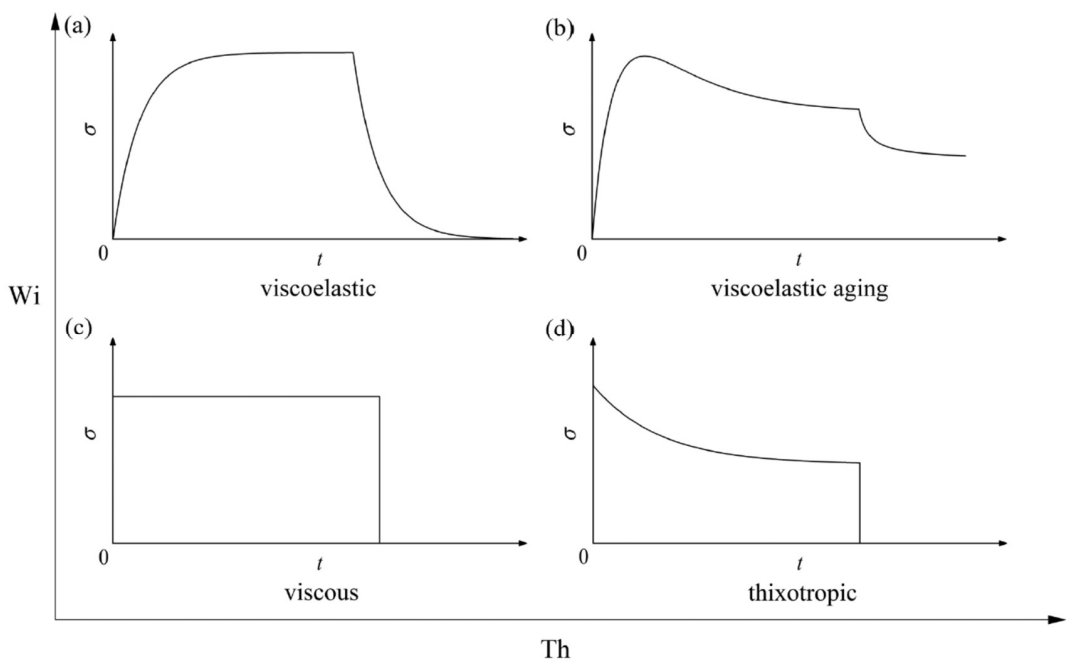

**Fig. 2** Idealized categorization proposed by Larson and Wei [3] of (a) viscoelastic, (b) viscoelastic aging, (c) viscous, and (d) thixotropic responses during shear start-up (instantaneous step jump in $\dot{\gamma}$ to $\dot{\gamma}_0$ at $t = 0$) succeeded by a stress relaxation ($\dot{\gamma}$ instantaneously jumps to 0). The overall vertical axis shows Weissenberg number ($Wi$, a product of shear rate and viscoelastic relaxation

time), while the horizontal axis shows thixotropy number ($Th$, a product of shear rate and thixotropic timescale). (Reprinted with permission from Ronald G. Larson and Yufei Wei, Journal of Rheology 63, 477-501 (2019), Copyright 2019, The Society of Rheology)

A similar ambiguity is present in the definitions of viscoelasticity and viscoelastic aging. It is not an independent topic as a significant section of the literature considers viscoelastic aging/rejuvenating materials to be thixotropic [2, 9]. Recently, Larson and coworkers [3] proposed a criterion to distinguish between viscoelasticity and viscoelastic aging by representing different stress responses in a shear start-up experiment followed by shear cessation. As shown in Fig. 2, they characterized the rheological responses as (a) viscoelastic, (b) viscoelastic aging, (c) viscous, and (d) thixotropic. For a purely viscous material (inelastic non-thixotropic), the stress achieves a constant value upon start-up and reduces to zero upon removal of shear rate, both instantaneously (Fig. 2(c)). Furthermore, since Larson and coworkers [3] describe (ideal) thixotropy as a purely viscous (inelastic) phenomenon, after applying a step-up jump in the shear rate, the stress instantaneously reaches a high value and then decreases to reach a steady-state. After cessation of flow, stress immediately goes to zero, as shown in Fig. 2(d). In a viscoelastic material, the stress reaches a plateau subsequent to start-up and shows a time-dependent relaxation of the same after the shear rate is reduced to zero (Fig. 2(a)). Furthermore, Larson and coworkers [3] propose that viscoelastic aging materials show a stress overshoot followed by an attainment of a steady constant value, as also observed for viscoelastic (non-aging) material. However, as per the proposal, after the cessation of flow, the stress gradually relaxes and may attain a plateau with a finite stress (residual stress), as shown in Fig. 2(b). Interestingly, for a time evolving material that attains the equilibrium state, stress eventually relaxes to zero. The proposal, therefore, suggests that a key difference between viscoelastic (non-aging) material and viscoelastic aging material (including inelastic thixotropic) is the presence of a stress overshoot in the latter. It needs to be ascertained whether all the viscoelastic aging materials and models show such behavior.

The objective of this work is to examine various experimental protocols meant to assess thixotropy in comparison to viscoelasticity. To this end, we employ 1.5 weight % aqueous PEO solution and also model the phenomena using the FENE-P model. Both these systems have been universally accepted as viscoelastic (non-thixotropic) in the literature [29, 47-49]. We also consider two model time dependent systems, namely aqueous clay dispersion and viscoelastic aging model proposed by Joshi [50]. According to Larson and coworkers [29, 51], both these systems do not qualify to be called thixotropic over much of the timescale of interest, as they do not reach equilibrium and show physical aging/rejuvenation dynamics. However, most papers in the literature consider aqueous clay dispersion that shows structural evolution as a function of time under quiescent conditions and time dependent decrease in viscosity under deformation field to be thixotropic over all the observation timescales [52-57]. We also adopt the same terminology and call both the clay dispersion and viscoelastic aging model to be thixotropic irrespective of time elapsed since shear melting. We subject these four systems to the various protocols available in the literature to assess whether viscoelastic response can be distinguished from the thixotropic response, and vice versa.

The outline of the paper is as follows. We discuss materials used and the experimental protocol in section II. The constitutive models and the governing equations are described in section III. We assess the following conventional protocols used for distinguishing thixotropy from viscoelasticity: step down jump in shear rate in subsection IV.A and shear startup flow at different waiting times in subsection IV.B. We also discuss stress relaxation and creep flow at different waiting times in subsection IV.C. In subsection IV.D we analyze viscoelastic aging behavior, both experimentally and theoretically. Finally, we end with summary and concluding remarks in section V.

## II. Material, sample preparation and experimental protocol

In this work, we investigate three types of materials: an aqueous solution of Polyethylene oxide, an aqueous dispersion of LAPONITE RD® clay soon after mixing clay with water, and an aqueous dispersion of LAPONITE RD® clay and salt. Polyethylene oxide (PEO) having molecular weight MW $8\times10^6$ gm/mol used in this study is purchased from Polysciences, Inc. The aqueous solution of PEO is prepared by dissolving the required amount of polymer (1.5 wt.%) in ultrapure water (resistivity 18.2 MΩ.cm). The solution is stirred gently in a magnetic stirrer for 12 hours at room temperature (25°C). LAPONITE RD® used in this work is procured from BYK Additives Inc. LAPONITE RD® is a synthetic clay system belonging to a family of hectorite and is composed of a disk-shaped particle having 25-30 nm diameter and 1 nm thickness [58]. This system hereafter will be referred to as clay. The aqueous dispersion of the same is prepared by dissolving the required amount of oven-dried (120 °C for four hours) clay and NaCl (if any) in ultrapure water. The sample is stirred for 30 minutes using Ultra Turrax drive at room temperature. In this work, we use two types of clay dispersions. The first type is 3.5 wt. % dispersion without any salt, 407 days after its preparation. In this case, the dispersion was preserved in an airtight polypropylene container for the whole duration, and all the experiments were performed within the next 20 days using the 407 days old sample. The second type is 2.67 wt. % clay with 3mM NaCl, which was used immediately after the stirring is over.

In this study, all the rheological experiments are performed using Anton Paar MCR 501 Rheometer. We use concentric cylindrical geometry with serrated walls (CC27/S, part no. 7108) with outside and inside diameters of 28.915 mm and 26.65 mm, respectively. Serrated geometry is used to avoid any slippage at the wall. Prior to any experiment, we pre-shear the sample at a high shear rate to eliminate any shear history associated with the same. We preshear 3.5 wt.% clay dispersion at 50 s$^{-1}$ for 300 s and 1.5 wt. % PEO solution at 10 s$^{-1}$ for 60 s. In order to get the same initial condition, we monitor the viscosity of a sample during pre-shear. After pre-shear, we wait for 600 s by applying $\dot{\gamma}=0$ before subjecting it to any deformation field, unless otherwise mentioned. In this work, we perform frequency sweep experiments, shear start-up experiments at different waiting times, stress relaxation and creep experiments at different waiting times, and step-down shear rate experiments. We cover the free surface of samples with a layer of low viscosity silicon oil, which prevents evaporation. All experiments are carried out at 25°C.

### III. Governing equations and constitutive models

In this work, we solve the simple shear flow for two constitutive models between two parallel plates separated by a distance $H$ in $y^*$ direction. The top plate is driven by a constant velocity $U$ in $x^*$ direction, whereas $z^*$ is the vorticity direction. The two plates are considered to extend infinitely in $x^*$ and $z^*$ direction. We assume the fluid to be incompressible. Hence, in the absence of inertia, the conservation of mass gets satisfied by itself for simple shear flow. The Cauchy momentum equation is then given by:

$$\nabla \cdot \underline{\underline{\sigma}}^* = 0, \qquad (1)$$

where, $\underline{\underline{\sigma}}^*$ is the deviatoric stress tensor. We use the following notation in this manuscript for all the variables: '*' superscript for dimensional (otherwise non-dimensional), '$i$' subscript for step-up shear rate flow, '$e$' subscript for step-down shear rate flow, and '$w$' subscript for the waiting period (For the modulus and compliance, we do not use superscript '*' to denote dimensional parameters, instead we represent it by $\bar{G}$ and $\bar{J}$, respectively.). For all the flow protocols considered herein, no-slip boundary condition is assumed. Of the two constitutive equations, one is an equilibrium viscoelastic model, while the other is a viscoelastic aging model. The governing equations, non-dimensionalization scheme for the respective constitutive models studied are as follows.

Conventional (non-thixotropic) viscoelastic models such as linear Maxwell model, upper convected Maxwell models, Oldroyd A/B models are linear or quasi-linear models and hence cannot be used to distinguish between thixotropy and non-linear viscoelasticity. On the other hand, generalized Newtonian (inelastic) models like power law and Carreau-Yasuda model that can predict shear thinning (a non-linear effect) are also not sufficient for the present purposes owing to lack of elasticity. For the purposes of this work, any non-linear viscoelastic model that shows shear thinning can be used for the intended purpose. The equilibrium viscoelastic constitutive equation considered here is the FENE-P (Finitely Extensible Nonlinear Elastic - Peterlin) model. This is a non-linear model, which was first introduced as FENE model by Warner [59]. FENE model considers polymer chains as dumbbells connected with a finitely extensible spring in contrast to a Hookean spring assumed in the Oldroyd-B model. But to allow for a closure in the constitutive relation of the FENE model, the Peterlin approximation is used, and the combination is referred to as the FENE-P model [60]. This viscoelastic model demonstrates non-linear effects such as shear thinning and first normal stress differences in a polymeric solution [61, 62]. The dimensional form of the constitutive equation for the total stress [60], which is assumed to be a sum of the solvent and polymeric contributions, is given by:

$$\underset{\sim}{\sigma}^* = -\eta_s \underset{\sim}{\dot{\gamma}}^* + \underset{\sim}{\sigma}_p^*, \tag{2}$$

where, $\underset{\sim}{\sigma}_p^*$ is polymeric stress contribution to the total deviatoric stress and $\underset{\sim}{\dot{\gamma}}^*$ is the rate of strain tensor. The polymeric stress is governed by the constitutive equation as follows:

$$Z\underset{\sim}{\sigma}_p^* + \tau^* \overset{\nabla}{\underset{\sim}{\sigma}_p^*} - \tau^* \left( \underset{\sim}{\sigma}_p^* - (1-\varepsilon b)\frac{\eta_p \underset{\sim}{\delta}}{\tau^*} \right) \frac{\mathrm{D}\ln Z}{\mathrm{D}t^*} = -(1-\varepsilon b)\eta_p \underset{\sim}{\dot{\gamma}}^*, \tag{3}$$

where, $\varepsilon = \dfrac{2}{b(b+2)}$, $Z = 1 + \left(\dfrac{3}{b}\right)\left(1 - \tau^* \dfrac{\mathrm{tr}\underset{\sim}{\sigma}_p^*}{3\eta_p}\right)$, $\underset{\sim}{\delta}$ is identity tensor, $\mathrm{tr}\underset{\sim}{\sigma}_p^*$ is the trace of $\underset{\sim}{\sigma}_p^*$ while $b$ is the dimensionless parameter denoting square of maximum (finite) extensibility of a polymer chain [63]. Here, the scalar $\dot{\gamma}^*$ is the second invariant of $\underset{\sim}{\dot{\gamma}}^*$ and in the present scheme, it is always $U/H$.

The non-dimensionalization scheme used for FENE-P model is as follows: $\underset{\sim}{\sigma} = \dfrac{\underset{\sim}{\sigma}^*}{((\eta_s + \eta_p)/\tau)}$, $\underset{\sim}{u} = \dfrac{u^*}{U}$, $y = \dfrac{y^*}{H}$, $t = \dfrac{t^*}{\tau^*}$, $\beta = \dfrac{\eta_s}{\eta_s + \eta_p}$, where, $\eta_s$ and $\eta_p$ are the contributions to the zero shear viscosity from the solvent and polymer respectively, $\underset{\sim}{u}$ is the velocity, $t$ is the time, and $\tau^*$ is the relaxation time. The Weissenberg number (dimensionless shear rate) is given by: $Wi = \dfrac{\tau^* U}{H}$. We solve FENE-P model using Bird et al.'s [60] sign convention for stress. However, while plotting the same, we use the standard sign convention.

We also study the viscoelastic aging constitutive equation proposed by Joshi [50]. Physical aging is known to decrease flowability of a material – also represented as fluidity ($\phi$) – while application of deformation field is known to enhance the same. The model proposed by Joshi considers the rate of change of $\phi$ as a first-order process. In the evolution equation, $\phi$ decreases with time ($t^*$) and increases with magnitude of stress according to [50]:

$$\frac{d\phi}{dt^*} = -\frac{\phi}{\tau^*(\phi)} + (1-\phi)\frac{\sigma^*}{\tau^*(\phi)\bar{G}(\phi)}, \tag{4}$$

where $\sigma^*$ is the second invariant of the stress tensor $\underset{\sim}{\sigma}^*$ and hence is always positive. The relaxation time $\tau^*$ and modulus $\bar{G}$ depend only on the instantaneous value of $\phi$ given by:

$$\tau^*(\phi) = \tau_0^* \left(1 + (\mu-1)A\ln\phi\right)^{\mu/(1-\mu)}, \text{ and} \tag{5}$$

$$\bar{G} = \bar{G}_0 \left(1 + B\ln(\tau^*/\tau_0^*)\right) \tag{6}$$

where $\mu$, $A$ and $B$ are the model parameters. The case of constant modulus is represented by $B=0$. The fluidity is normalized in such a fashion that $\phi=1$ represents the completely rejuvenated or shear melted state. while $\phi=0$ represents the state with minimum possible fluidity. In Eqs. (5) and (6), $\tau_0^*$ and $\bar{G}_0$ are respectively the relaxation time and modulus of a material at $\phi=1$. A time-dependent Maxwell model is used as the constitutive equation:

$$\underset{\sim}{\dot{\gamma}}^* = \frac{\underset{\sim}{\sigma}^*}{\bar{G}(\phi)\tau^*(\phi)} + \frac{d}{dt^*}\left(\frac{\underset{\sim}{\sigma}^*}{\bar{G}(\phi)}\right), \tag{7}$$

The physics associated with specific forms used in Eq. (5) and (6) is discussed in [50]. We use the following non-dimensional parameters: $\sigma = \sigma^*/\bar{G}_0$, $G = \bar{G}/\bar{G}_0$, $t = t^*/\tau_0^*$, $\tau = \tau^*/\tau_0^*$ and $\text{Wi} = \tau_0^*\dot{\gamma}^*$, where $\dot{\gamma}^*$ is the second invariant of the rate of strain tensor $\underset{\sim}{\dot{\gamma}}^*$. While solving the model, we always consider the initial state ($t=0$) to be the completely rejuvenated state given by $\phi=1$ and $\bar{G}_0$ is the corresponding modulus. Under quiescent conditions, the model predicts that relaxation time to have a power law dependence on time given by: $\tau \sim t^\mu$ as routinely observed experimentally [42, 43, 46, 64-66]. Here $\mu<1$, $\mu=1$, and $\mu>1$ respective represents sub-aging, full-aging and hyper-aging regimes [67, 68].

For a specified flow protocol, the constitutive model along with the Cauchy momentum equation is solved using the ode45 package of MATLAB®. The ode45 package uses the explicit

Runge-Kutta (4,5) method. In the case of viscoelastic model, the results from the FENE-P model are verified with the analytical solution of the Oldroyd-B model in the limit $b \to \infty$. For the viscoelastic aging model, our results were verified with the results published in [50].

**IV. Results and discussion**

The two experimental systems that we investigate are: 407 days old 3.5 wt.% aqueous clay dispersion, a model time dependent material, and 1.5 wt.% aqueous PEO solution, an equilibrium viscoelastic material. We first characterize these materials by performing a frequency sweep experiment at $\gamma = 1\%$ (linear response regime) subsequent to waiting for 600 s after pre-shearing the samples. During the waiting period, we subject the sample to $\dot{\gamma} = 0$. In Fig 3(a), the elastic ($G'$) and viscous ($G''$) moduli are plotted as a function of the angular frequency ($\omega^*$) for 3.5 wt.% clay dispersion. We only explore sufficiently high frequencies to prevent physical aging of the material over a course of single oscillation. It can be seen that both the moduli remain practically constant over the explored frequencies. In Fig. 3(b), we plot $G'$ and $G''$ as a function of $\omega^*$ for 1.5 wt.% PEO solution. Since the PEO solution is an equilibrium system, we have explored the frequencies over a broad range. The system shows a terminal behavior in the limit of low frequencies. With an increase in frequency, eventually, both the moduli cross over ($G' = G''$). A characteristic relaxation time of the PEO solution is obtained by taking the inverse of frequency at which the crossover occurs ($G' = G''$) that can be estimated to be around 10.3 s. This suggests that after removal of any deformation field, PEO solution should reach the equilibrium state, when kept under quiescent conditions, after a time of the order of 10 s has elapsed. We also obtain the steady-state flow curves for both the systems. For clay dispersion, the steady-state flow curve obtained after shearing the sample at respective shear rates for a long time is plotted in Fig 3(c). It can be seen that the dispersion shows yield stress to be around 36 Pa. In the inset of figure 3 (c), we also plot the steady-state flow curve for 1.5 wt.% PEO solution, which shows the Newtonian behavior below the shear rate of 0.1 s$^{-1}$.

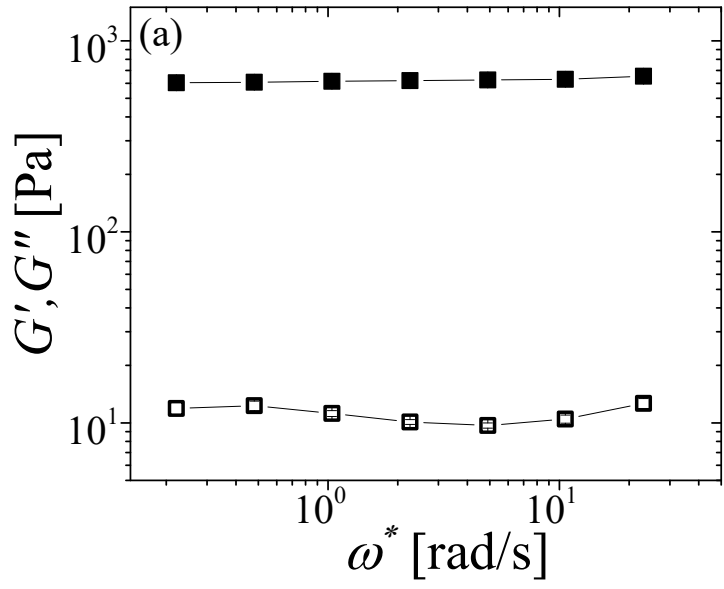

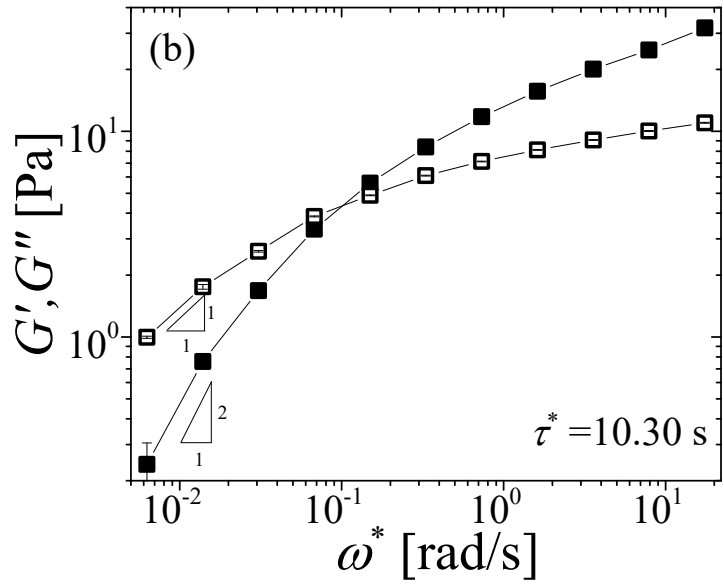

$\tau^* = 10.30$ s

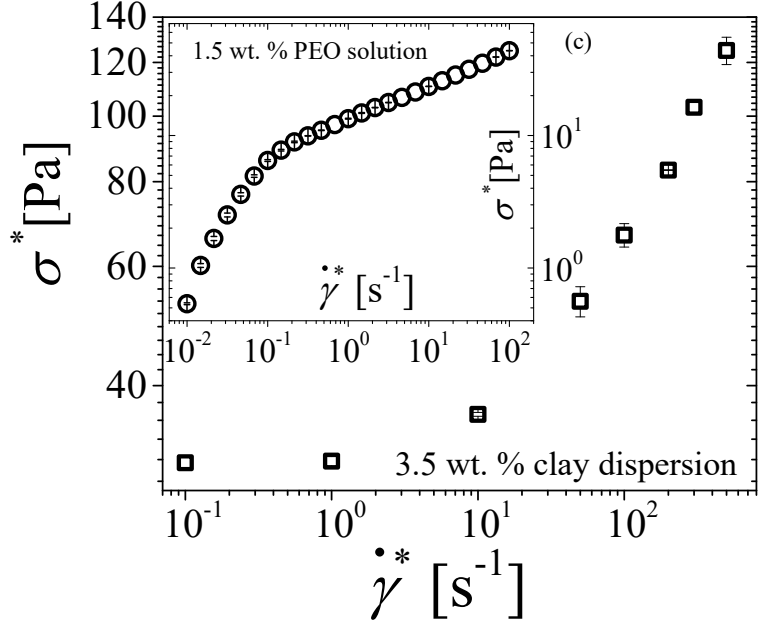

**Fig. 3.** The elastic ($G'$, closed symbols) and viscous ($G''$, open symbols) moduli are plotted as a function of the angular frequency $(\omega^*)$ for 3.5 wt.% clay dispersion in (a) and 1.5 wt.% PEO solution in (b). The frequency sweep experiment is carried out at strain amplitude $\gamma=1\%$ and waiting time of 600 s (time elapsed since pre-shear). The steady-state flow curve of 3.5 % clay dispersion is plotted in (c) while the steady-state flow curve of 1.5 % PEO solution is plotted in the inset of (c).

### A. Step-down jump in the shear rate

We now investigate both the systems by applying the distinguishing criterion shown in Fig. 1(a). According to the experimental protocol, we measure the stress/viscosity transients when a material is subjected to step-down in shear rate from $\dot{\gamma}_i^*$ to $\dot{\gamma}_e^*$. In the first protocol, a finite shear rate $\dot{\gamma}_i^* = 300$ s$^{-1}$ (for clay dispersion) is applied for 300 s, which allows the system to attain a steady-state. Subsequently, the shear rate is dropped from $\dot{\gamma}_i^*$ to $\dot{\gamma}_e^* = 10, 1, 0.1$ s$^{-1}$. In Fig 4(a), the corresponding stress evolution is plotted for 3.5 wt.% clay dispersion. For a given $\dot{\gamma}_i^*$, the nature of stress transient after the step down in shear rate depends on the value of $\dot{\gamma}_e^*$. For the largest explored value of $\dot{\gamma}_e^* = 10$ s$^{-1}$, the stress undergoes a step decrease from a value associated with $\dot{\gamma}_i^*$. The corresponding stress value is shown by an arrow on the left vertical axis. For time $t^* > 0.1$ s, shear stress evolution associated with $\dot{\gamma}_e^* = 10$ s$^{-1}$ shows a weak increase as a function of time. With the decrease in $\dot{\gamma}_e^*$ below 10 s$^{-1}$, the stress decreases gradually after the step-down jump in shear rate, shows a minimum, undergoes an increase and attains a steady-state plateau. Furthermore, the value of the stress minimum can be seen to be decreasing. In Fig 4(b), the stress transient is plotted for a protocol, wherein the shear rate is dropped from different $\dot{\gamma}_i^* = 500, 300, 100$ s$^{-1}$ to $\dot{\gamma}_e^* = 1$ s$^{-1}$. In all the cases, after the down jump, the stress decreases gradually, and after reaching a minimum, increases and attains a plateau at steady-state. For this protocol, the stress at minimum decreases with an increase in $\dot{\gamma}_i^*$ (or increase in $\dot{\gamma}_i^* - \dot{\gamma}_e^*$). The behavior shown in Fig. 4(a) and (b), therefore, corroborates with the trend shown in Fig 1(d), suggesting that the aqueous clay dispersion studied is a viscoelastic thixotropic material.

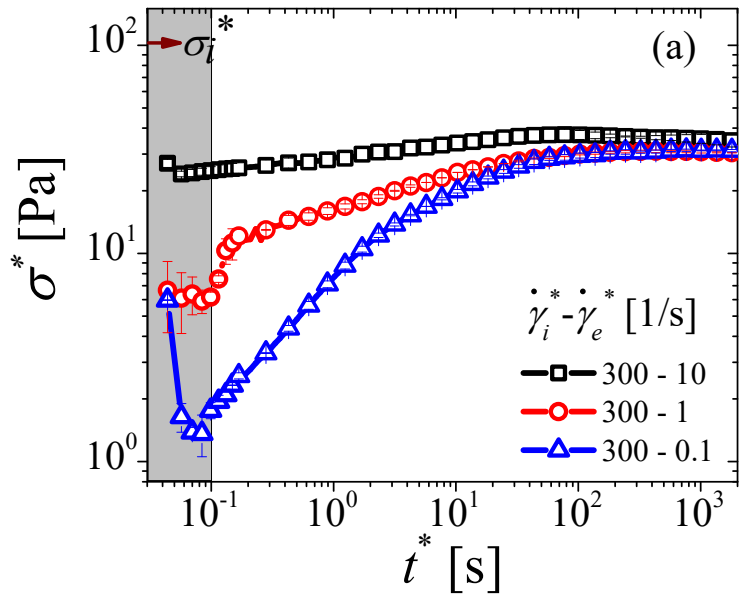

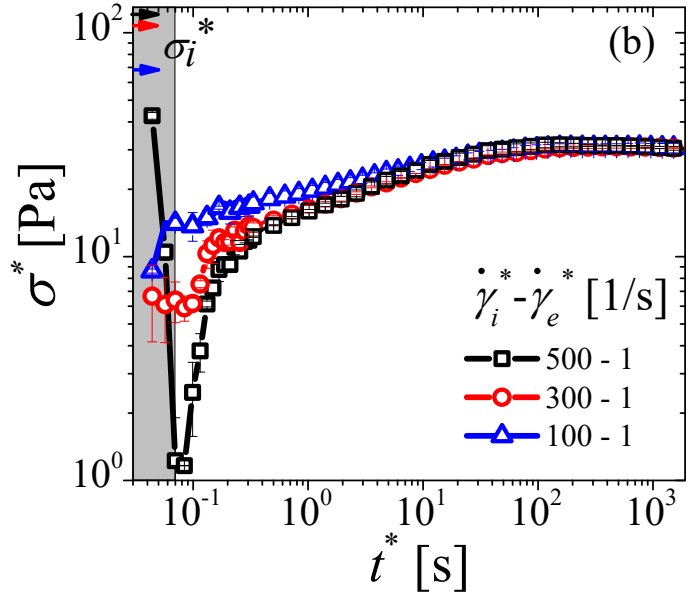

**Fig. 4**. Transient evolution of stress, resulting from a step-down jump in the shear rate is plotted for the two cases: (a) from $\dot{\gamma}_i^* = 300$ s$^{-1}$ to $\dot{\gamma}_e^* = 10, 1, 0.1$ s$^{-1}$ and (b) from $\dot{\gamma}_i^* = 500, 300, 100$ s$^{-1}$ to $\dot{\gamma}_e^* = 1$ s$^{-1}$, for 3.5 wt.% clay dispersion. The arrow on the left vertical axis corresponds to the stress value just before the application of $\dot{\gamma}_e^*$. The data below 0.07 s is shown by gray background as it is affected by instrument inertia.

In Fig. 5 (a), transient in stress after step-down in shear rate has been plotted for 1.5 wt. % PEO solution with $\dot{\gamma}_i^* = 100$ s$^{-1}$ and $\dot{\gamma}_e^* = 1$ to $0.001$ s$^{-1}$. After step-down, the stress decreases gradually and, depending upon the value of $\dot{\gamma}_e^*$, it either reaches a steady value; or shows a minimum, increases and then eventually reaches a constant steady value. The value of stress at the point of minimum decreases (a result akin to Fig. 4(a) results of clay dispersion) while the time at which the stress minimum occurs increases with decrease in $\dot{\gamma}_e^*$. We also perform the shear rate step-down experiments from different $\dot{\gamma}_i^* = 100$ to $1$ s$^{-1}$ to $\dot{\gamma}_e^* = 0.1$ s$^{-1}$. The transient stress shows a gradual decrease with time, attains a minimum and then increases again with time and achieves a steady-state plateau. Similar to that observed for the clay dispersion, the value of stress associated with the minimum decreases with increase in $\dot{\gamma}_i^*$ (or increase in $\dot{\gamma}_i^* - \dot{\gamma}_e^*$). The time at which the minimum occurs, however, increases weakly with decrease in $\dot{\gamma}_i^*$. The behavior similar to Fig 5(a), has also been reported for entangled polybutadiene solution, a well-known viscoelastic system, by Wang and coworkers [69]. Interestingly, the behavior shown by PEO solution and polybutadiene solution also corroborates with the qualitative trend shown in Fig. 1(d), and therefore would lead one to conclude these systems to be (viscoelastic) thixotropic. However, these systems are well accepted to be non-thixotropic viscoelastic.

In order to verify whether the experimental data in Figs. 4 and 5 is not affected by the instrument inertia, we monitored the shear rate values after the step change in the shear rate. We observe that it takes around 0.1 s to achieve the final shear rate after a step down for the clay dispersion, the PEO solution as well as for a silicone oil. The high shear rate viscosity of 3.5 wt. % clay dispersion and silicone oil are of same order of magnitude (0.3-0.5 Pa s). This suggests that the experimental data in Figs. 4 and 5 only below 0.1 s is affected by the instrument inertia. However, it does not affect the interpretation of the clay dispersion results as it anyway expected to show an undershoot in stress, and the undershoot continues well beyond 0.1 s. For the PEO solution, inertia affects the data below $t < 10^{-2}$ ($t^* = 10.3 t < 0.1$ s) which is well beyond the range plotted in Fig 5. Therefore, the data presented for PEO solution is not affected by the instrument inertia.

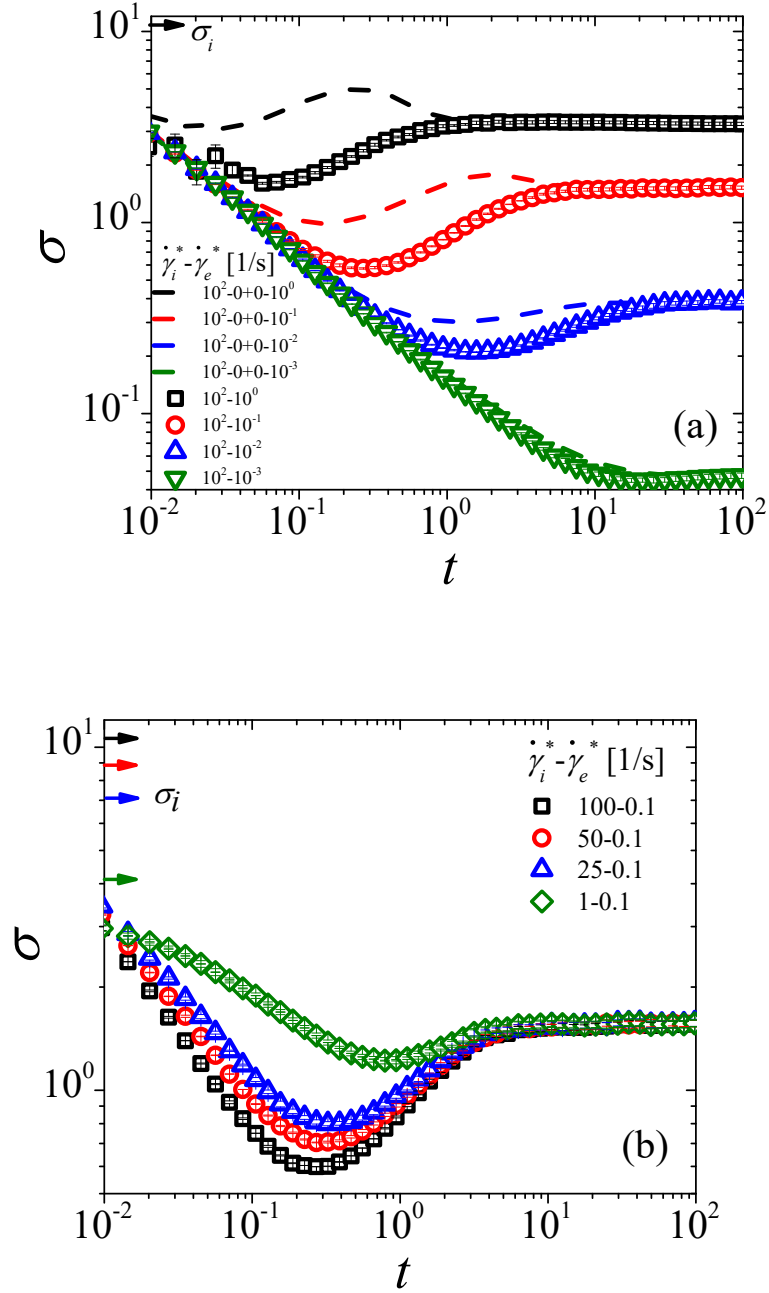

**Fig 5.** Transient evolution of dimensionless stress, resulting from a step-down jump in the shear rate is plotted for the two cases: (a) from $\dot{\gamma}_i^* = 100$ s$^{-1}$ to $\dot{\gamma}_e^* = 1, 0.1, 0.01, 0.001$ s$^{-1}$ and (b) from $\dot{\gamma}_i^* = 100, 50, 25, 1$ s$^{-1}$ to $\dot{\gamma}_e^* = 0.1$ s$^{-1}$, for 1.5 wt. % aqueous PEO solution. In figure (a), the dashed line shows a linear superposition of stress $\left[\sigma(\dot{\gamma}_i - 0) + \sigma(0 - \dot{\gamma}_e)\right]$ obtained from the cessation of steady shear flow $(\dot{\gamma}_i - 0)$ and start-up shear flow $(0 - \dot{\gamma}_e)$. Shear stress is non-dimensionalised using $G'$ (or $G''$) value at crossover point in the frequency sweep plot of the solution (4.36 Pa), and time units are non-dimensionalised using relaxation time of the solution (10.3 s). The arrow on the left vertical axis corresponds to the stress value just before the application of $\dot{\gamma}_e^*$.

We next discuss the results obtained using the FENE-P model, which was subjected to the same step-down shear rate protocol, from $Wi_i$ to $Wi_e$. We numerically solve Eqs. (1)-(3) to obtain the variation of stress as a function of time. The corresponding stress transients for step-down in shear rate from $Wi_i = 10^2$ to $Wi_e = 10$ to $10^{-3}$ are plotted (solid lines) in Fig. 6(a). We also perform simulations from different values of high shear rate over the range, $Wi_i = 10^7$ to $10^4$ to a single value of $Wi_e = 10^{-1}$, and the corresponding stress evolution is plotted in Fig. 6(b). It can be seen that during the evolution, the stress indeed exhibits a minimum, before increasing and reaching a constant steady-state value. This observation is qualitatively similar to experimental results of 1.5 wt.% PEO solution as well as the trend mentioned in Fig. 1(d) associated with the stress transient in a viscoelastic thixotropic material [9], which is qualitatively similar to what was reported for 2.9 vol.% fumed silica suspension for a step-down shear rate experiment [70].

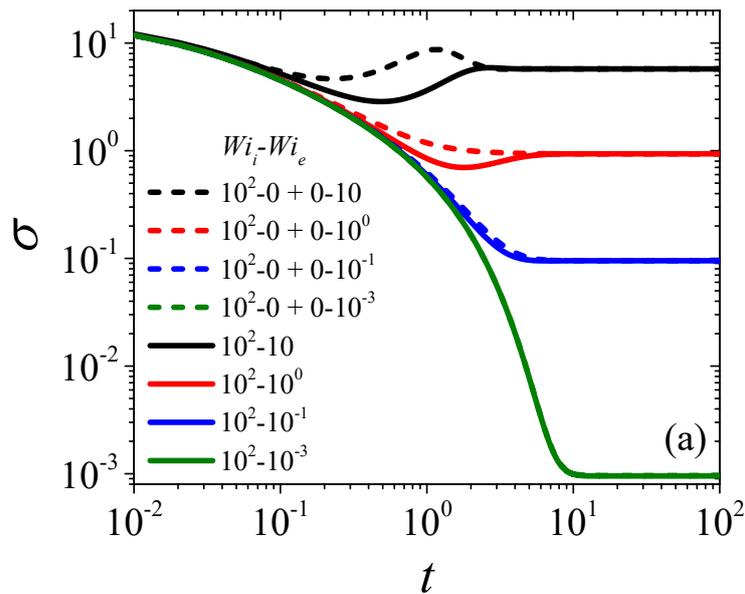

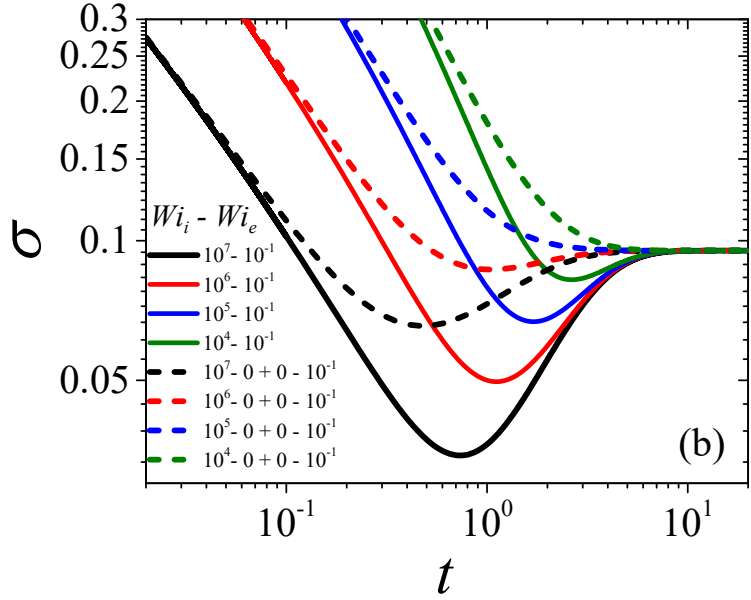

**Fig 6**. Transient evolution of stress, resulting from a step-down jump in the shear rate is plotted for the two cases: (a) from $Wi_i = 10^2$ to $Wi_e = 10^1$, $10^0$, $10^{-1}$, $10^{-3}$ and (b) from $Wi_i = 10^7$, $10^6$, $10^5$, $10^4$ to $Wi_e = 10^{-1}$ for FENE-P model with $b = 10^2$ and $\beta = 10^{-3}$. In both figures, the dashed line shows a linear superposition of stress $\left[\sigma(Wi_i - 0)) + \sigma(0 - Wi_e)\right]$ obtained from the cessation of steady shear flow $(Wi_i - 0)$ and start-up shear flow $(0 - Wi_e)$. The steady-state value of $\sigma$ during shear startup at $Wi = 10^2$, $10^4$, $10^5$, $10^6$, and $10^7$ respectively is 15.87, 78.03, 168.49, 363.16, and 782.49.

With respect to the behavior of 1.5 wt.% aqueous PEO solution and the results from the FENE-P model shown in Figs. 5 and 6, the step-down shear-rate protocol shows two kinds of stress transients, both with and without the stress undershoot. In both experimental observations and model predictions, the intensity of undershoot gets enhanced by increasing the value of initial shear rate ($\dot{\gamma}_i^*$). The effect of the magnitude of shear rate on the relaxation of stress subsequent to cessation of steady shear flow of polymeric solutions has been studied by Menezes and Graessley [71]. They observed that the higher the shear rate (also implies higher corresponding stress), higher is the initial rate of relaxation of stress. (For the Oldroyd B model, since the relaxation rate is independent of the shear rate that results in constant viscosity, an undershoot in stress evolution for the studied flow field is never observed). This is because, in a step jump experiment, the immediate response of the material would correspond to the rheological properties of the initial state. Eventually, the response of the material becomes a function of its properties at its final state [71-73]. Consequently, the greater the $\dot{\gamma}_i^*$ (or $Wi_i$), lesser is the time in which the stress decreases. However, if $\dot{\gamma}_e^*$ (or $Wi_e$) is sufficiently large, so that the steady-state stress associated with the same is greater than the initial stress decrease, the system must show a stress undershoot. Nevertheless, if the steady-state stress associated with $\dot{\gamma}_e^*$ (or $Wi_e$) is lower compared to the initial decrease in stress, then the stress reaches an ultimate plateau monotonically without showing an undershoot. The observed experimental behavior is also consistent with this scenario. In order to contrast the effect of non-linear viscoelastic response from the linear viscoelastic response on the observed behavior, we plot the linear superposition of two shear flows ($\dot{\gamma}_i^*$ (or $Wi_i$) to 0 superposed on 0 to $\dot{\gamma}_e^*$ (or $Wi_e$)), wherein the former is the cessation of steady shear flow ($\dot{\gamma}_i^*$ (or $Wi_i$) to 0) while the latter is the start-up shear flow (0 to $\dot{\gamma}_e^*$ (or $Wi_e$)), and are represented by the dashed lines in Figs. 5(a), 6(a) and 6(b). It is known that in the limit of linear viscoelasticity, the response to an independent application of the flow field is additive. Interestingly, the dashed lines in Figs. 5(a), 6(a) and 6(b) suggest that for the cases where the system shows a stress undershoot, the linear superposition behavior is at variance from the shear rate step-down stress evolution data. On the other hand, when the stress response does not show undershoot, the linear combination of the two independent flow fields matches the stress evolution associated with the step-down shear rate response very well. In order to emphasize this aspect, we also plot the shear rate dependence of viscosity for 1.5 wt.% aqueous PEO solution and the FENE-P model in Fig. 7. It can be seen that if the steady-state viscosity associated with $Wi_i$ and $Wi_e$ is not same, then the corresponding step-down shear rate results in undershoot of stress. Overall, the above discussion clearly suggests that non-linear viscoelasticity along with shear thinning plays a crucial role in the appearance of an undershoot upon a step-down decrease in shear rate.

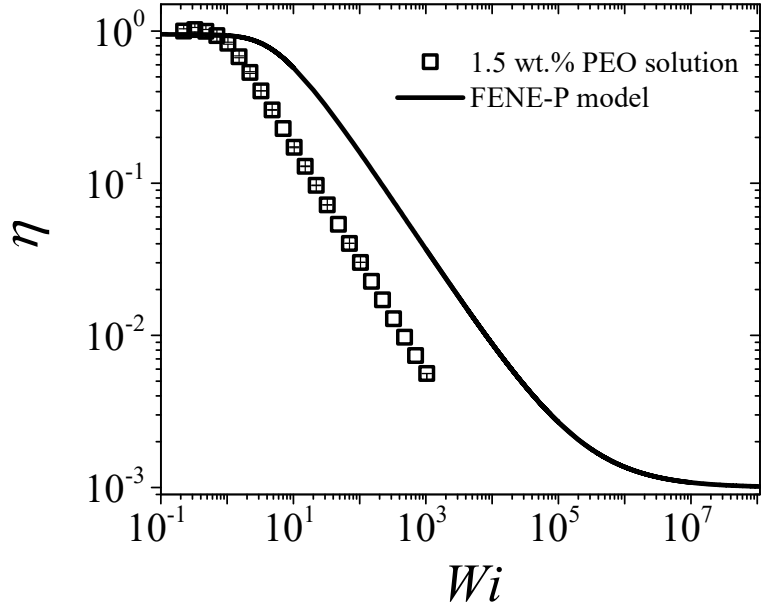

**Fig 7.** Dimensionless viscosity is plotted as a function of dimensionless shear rate (Weissenberg number, $Wi$) by varying the shear rate from $Wi = 10^{-1} - 10^8$ for FENE-P model ($b = 10^2, \beta = 10^{-3}$) (line) and 1.5 wt.% PEO solution (symbols). For PEO solution, the viscosity of 1.5 wt.% PEO solution is non-dimensionalised using zero shear viscosity of the solution (77.8 Pa s), and shear rate is non-dimensionalised using relaxation time of the solution (10.3 s)

In thixotropic viscoelastic materials, the origin of stress undershoot is attributed to two timescales. One is viscoelastic relaxation time, and it corresponds to decrease in stress immediately after step down in shear rate is imposed. The other independent timescale is associated with structure build-up (also termed as the thixotropic timescale), and it relates to the build-up of stress past minimum in stress. According to Larson and Wei [3], in a thixotropic viscoelastic material, relaxation time is much smaller than the thixotropic timescale. Interestingly, the results of Fig. 5 (1.5 wt.% PEO solution) and Fig. 6 (FENE-P model) show that the stress build-up occurs over around 1 order of magnitude time greater than time associated with the stress decay. The polybutadiene system used by Wang and coworkers [69] shows the stress build-up timescale to be over two orders of magnitude higher than the timescale associated with stress decay. Consequently, these viscoelastic systems seem to exhibit an apparent thixotropic timescale being significantly larger than the viscoelastic timescale. As we have discussed, this behavior is solely due to non-linear viscoelasticity along with shear thinning nature of a material resulting in shear rate dependent relaxation rate. In addition, physically aging soft glassy materials are known to show non-monotonic shear stress – shear rate flow curve [12]. Consequently, application of low shear rate may induce transient as well as steady-state shear banding in the same [74, 75]. The step-down shear rate jump criterion, however, assumes homogeneous flow field, which may lead to further complications in the stress response. The above results clearly show that undershoot in stress subsequent to step-down in shear rate cannot be used as a distinguishing criterion for thixotropic behavior.

### B. Shear start-up at different waiting times

One of the fingerprints underlying thixotropy is the variation of the intensity of stress overshoot during shear start-up as a function of rest (waiting) time elapsed since pre-shear [2]. Such behavior is also considered as a signature of physical aging in soft glassy materials [76]. In a typical experiment, a material, which is kept under quiescent conditions over a certain waiting time after the pre-shear of identical magnitude and duration, is subjected to a constant shear rate, and the corresponding evolution of stress is monitored as a function of time. In a variety of physically aging (thixotropic) materials as well, it has been observed that the magnitude of stress overshoot increases with an increase in waiting time at which the step shear rate has been applied [32, 34, 46, 76]. An illustration of such behavior has been shown for 3.5 wt.% clay dispersion in Fig. 8. After pre-shearing the same for 300 s at 50 s$^{-1}$ shear rate, the sample was kept under quiescent (rest) conditions by applying a zero shear rate. The sample was then subjected to a shear rate of 3 s$^{-1}$ at different waiting times. The corresponding evolution of stress for shear start-up at different waiting times is plotted in Fig. 8. It can be seen that the stress shows an overshoot, whose magnitude indeed increases with increase in waiting time at which the material is subjected to the step shear rate. We perform the same test on the PEO solution, wherein, subsequent to strong pre-shear at 10 s$^{-1}$ and for 60 s, it has been kept at rest for different waiting times ($t_w^*$=5, 10, 30, 60, 120, 900, 1800 s, in dimensionless form $t_w = 0.49$ to 174.76) and then subjected to a constant shear rate of 1 s$^{-1}$. In Fig. 9, we plot the evolution of dimensionless stress as a function of dimensionless time elapsed since the application of step shear rate for different dimensionless rest times. It can be seen that the magnitude of overshoot goes on increasing up to a rest time of 60 s, which is of the order of the relaxation time of polymer solution. For $t_w^* > 60$ s, however, the material is in a relaxed state and application of step shear rate results in identical stress overshoot that superposes on to each other. We also solve the FENE-P model for the start-up shear subjected to a pre-shear at $Wi_1$ and steady-state is allowed to establish. Consequently, the shear rate was set to 0 for different waiting times, $t_w$, and then the model is subjected to a shear rate $Wi_2$ for $t > t_w$. The corresponding evolution of stress at different $t_w$ is plotted as a function of $t - t_w$ in Fig. 10. We observe that the stress increases and attains a steady-state value after showing an overshoot. The transient in an initial increase in stress, before the overshoot, depends on the waiting time because we do not consider inertia in the theoretical formulation. We observe that the intensity of shear stress overshoot increases as a function of waiting time until $t_w < 4$ (for times below that of the order of magnitude of relaxation time). However, for waiting times significantly larger than relaxation times, the overshoot superposes on to each other as expected for systems under equilibrium.

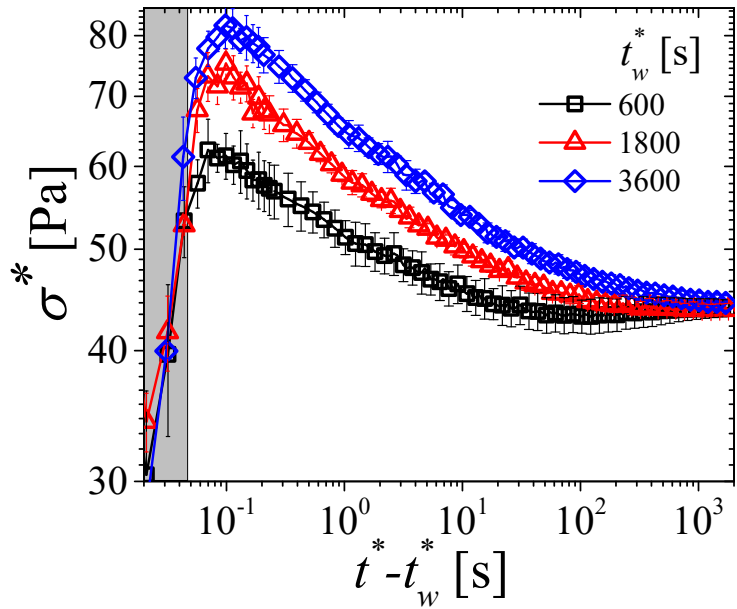

**Fig 8.** Stress transient $\sigma^*(t)$ is plotted against time for different waiting times $t_w^*$, as shown in legend for 3.5 wt.% clay dispersion for the applied shear rate of 3 s$^{-1}$. Shear stress overshoot increases with increase in the value of waiting time $t_w^*$. The data below 0.05 s is shown by gray background as it is affected by instrument inertia.

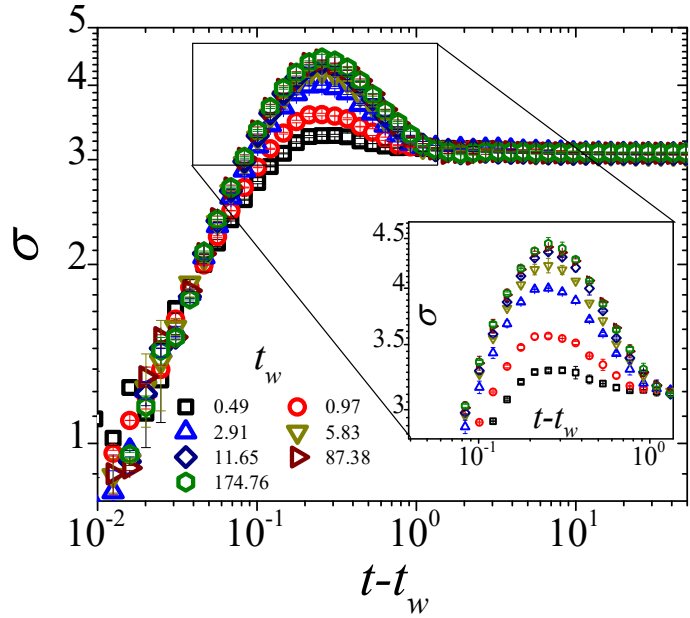

**Fig. 9.** Evolution of dimensionless stress $\sigma(t)$ is plotted as a function of dimensionless time for different dimensionless waiting times $t_w$, as shown in legend for 1.5 wt.% PEO solution for the applied shear rate of 1 s$^{-1}$. Shear stress is non-dimensionalised using $G'$ value at crossover point in the frequency sweep plot of the solution (4.36 Pa), and time units are non-dimensionalised using relaxation time of the solution (10.3 s).

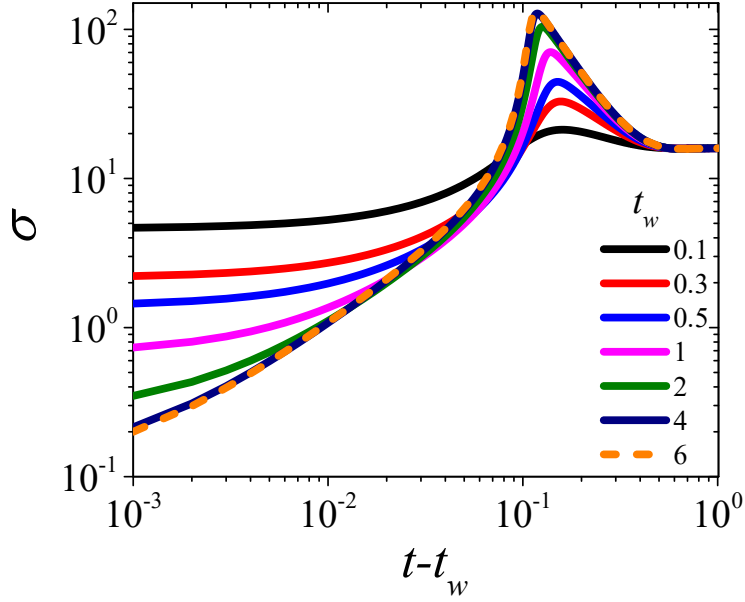

**Fig. 10**. Stress transient $\sigma(t)$ is plotted against time for different waiting times $t_w$ as shown in legend for FENE-P model $(b=10^2, \beta=10^{-3})$ pre-sheared at, $Wi_1 = 100$ and an applied shear rate, $Wi_2 = 100$. Shear stress overshoot height is sensitive to the value of $t_w$ in the domain when $t_w < \tau$.

The behavior shown by the 1.5 wt.% PEO solution and the FENE-P model indicate that when the rest time is below that of relaxation time of a material, the magnitude of stress overshoot depends upon the rest time and is qualitatively similar to what has been observed for the physically aging/thixotropic materials [34, 46]. It is known that the relaxation time of a viscoelastic material may span from nanoseconds to centuries [29]. Consequently, it can be concluded that, using the above protocol, it is not possible to distinguish a (equilibrium) viscoelastic material with a very large relaxation time from a thixotropic/physical aging material.

## C. Stress relaxation and Creep at different waiting times

Another experimental procedure often employed for the soft glassy materials to assess the physical aging behavior is the waiting time dependence of the creep compliance and/or relaxation modulus [43, 46]. In a typical protocol, similar to that used for start-up shear flow, a material is subjected to pre-shear until it reaches a steady-state (or dynamic steady-state, if it is subjected to oscillatory flow field of large amplitude). Subsequently, the material is kept under quiescent conditions to which step strain or step stress is applied at different waiting times. We perform both the step strain, and the step stress experiments-with respective strain and stress magnitudes in the linear viscoelastic region. The results of step strain experiments are shown below while the results of step stress experiments are presented in the supplementary information. For step strain (and step stress) experiments, upon cessation of pre-shear, during the quiescent period, we apply $\dot{\gamma}^* = 0$ so that the material undergoes stress relaxation. In Fig 11(a), we plot the evolution of stress after stopping the steady shear flow (50 s$^{-1}$ applied for 300 s) for the 3.5 wt.% clay dispersion. It can be seen that the stress relaxes over a timescale of the order of 10 s and shows a plateau of constant residual stress for higher times. We also apply step strain (1 %) to the system at different waiting times, and the subsequent relaxation of stress is plotted in terms of relaxation modulus in Figs. 11(b) and (c). For $t_w^* < 10$ s, the initial value of modulus decreases with an increase in $t_w^*$. Interestingly, the overall rate of relaxation of stress on the double logarithmic scale can also be seen to be decreasing with increase in $t_w^*$. Consequently, after a certain time, the relaxation modulus curves cross each other, as shown in Fig. 11(b). For $t_w^* > 10$ s, interestingly, the trend of the initial value of relaxation modulus reverses, and it increases with an increase in $t_w^*$ as shown in Fig. 11(c). Furthermore, the rate of relaxation also gets sluggish with increase in $t_w^*$. The results for creep experiments on 3.5% clay dispersion also show similar dependence of creep compliance on waiting time as shown in Fig. S1. At low waiting times, the initial value creep compliance increases with waiting times and then it starts to cross each other as shown in Fig. S1 (b). Figure S1 (c) shows that the value of creep compliance decreases with $t_w^*$ for higher values of $t_w^*$. The behavior similar to that shown in Figs. 11(c) and S1(c) is often reported for different soft glassy materials, wherein either relaxation of modulus gets slower (or increase in compliance becomes sluggish) with $t_w^*$ [43, 46, 65, 66, 77].

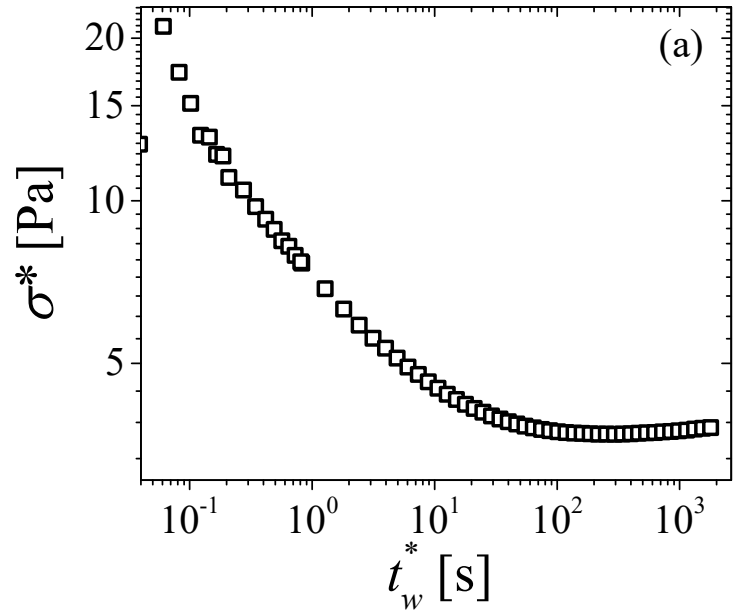

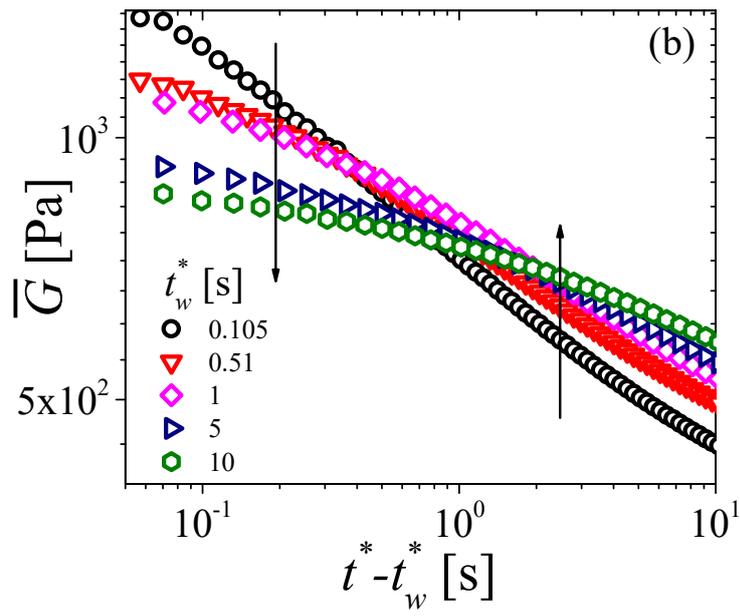

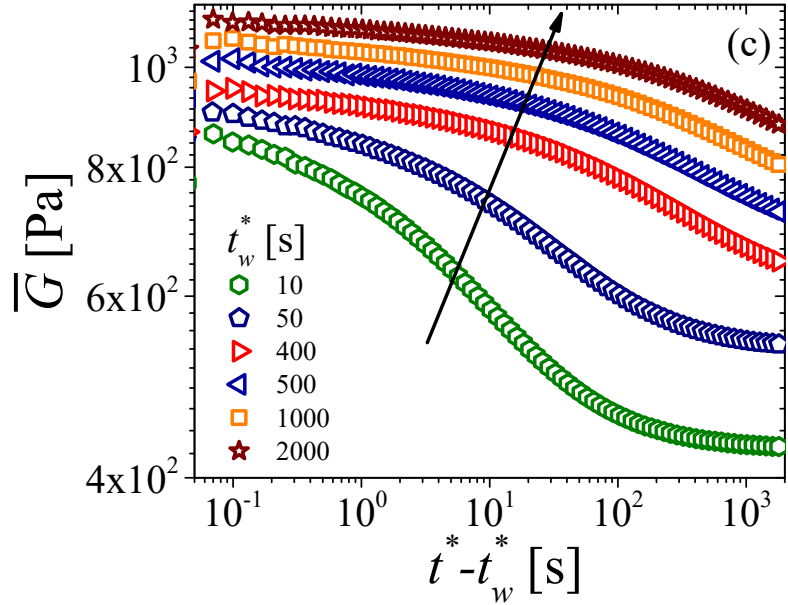

**Fig. 11.** (a) Shear stress relaxation subsequent to the cessation of steady shear flow at (50 s$^{-1}$ applied for 300 s) for 3.5 wt.% clay dispersion. It can be seen that the stress undergoes relaxation and reaches a plateau for $t_w^* > 10$ s. The sample is subjected to a step strain of 1 % at different waiting times after cessation of steady shear. The corresponding evolution of relaxation modulus is plotted for waiting times $\left(t_w^*\right)$ = 0.105 to 10 s (b) and = 10 to 2000 s (c). The arrows on the figure (b) and (c) indicates the direction of evolution of relaxation modulus $\left(\overline{G}\right)$ with increasing value of $t_w^*$ as a function of $t^* - t_w^*$.

We also solve the viscoelastic aging model with time dependent modulus [50] for $A=10$, $B = \ln 10$ and $\mu = 1.1$ subjected to a similar flow protocol. The corresponding relaxation of stress after the pre-shear is plotted in Fig. 12(a), which shows that the stress continuously goes on decreasing as a function of time (over the explored time duration), although the rate of decrease becomes very sluggish in a limit of large times. In response to the step strain over this period, the time evolution of stress relaxation modulus at different waiting times is plotted in Fig. 12(b) and (c). Importantly, the behavior of viscoelastic aging model is qualitatively similar to experimental results shown in Fig. 11(b) and (c). For $t_w < 70$, increase in $t_w$ leads to a decrease in the initial value of the relaxation modulus. The decay of the same is, however, faster for the smaller $t_w$; consequently, the relaxation modulus curves cross each other in accordance with the experimental data shown in Fig. 11(b). However, for $t_w > 70$, the initial relaxation modulus starts increasing and the relaxation becomes progressively slower with increase in $t_w$ as shown by the experimental data in Fig. 11(c). The results of viscoelastic aging model for creep experiments are shown in Fig. S2. The dependence of creep compliance on waiting time is equivalent to the dependence of relaxation modulus on waiting time. Figure S2 (b) and (c) shows that creep compliance at different $t_w$ but in a limit of low $t_w$ initially increase, crosses each other at later times. However, at higher $t_w$, compliance not just decreases with its evolution also get more retarded with increase in $t_w$.

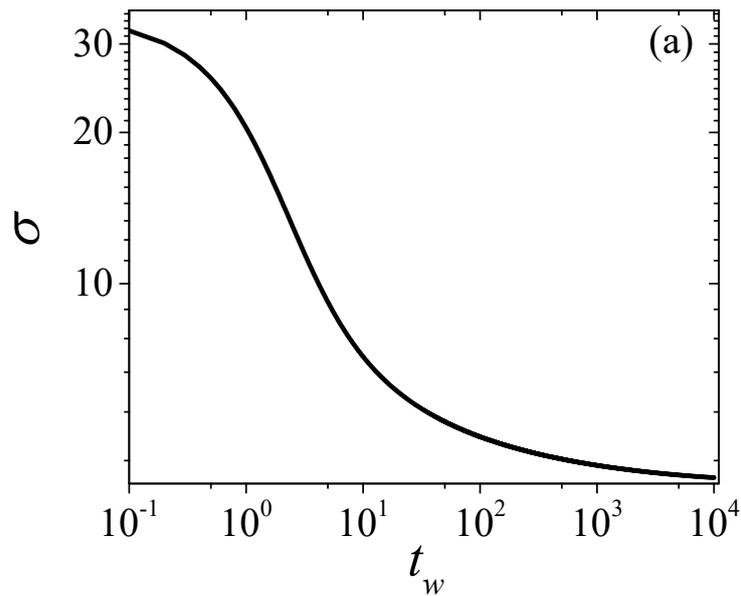

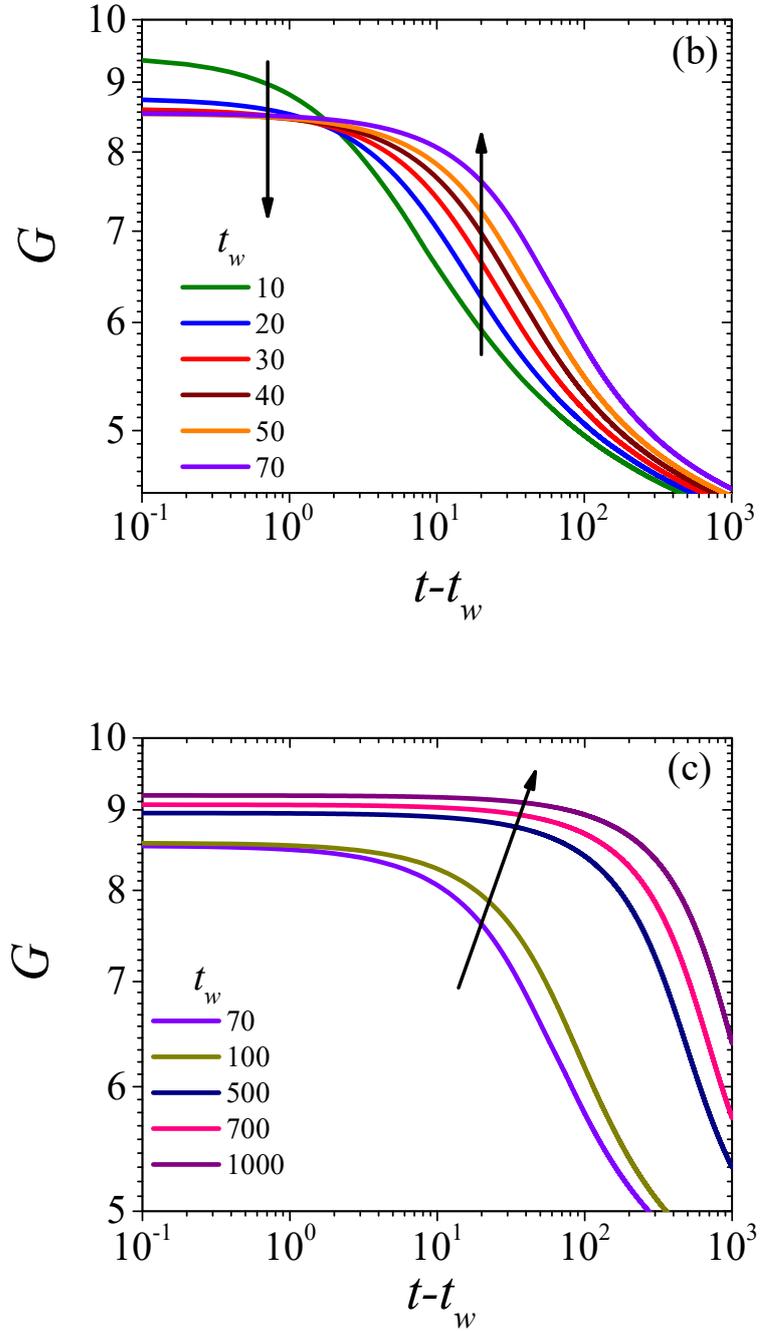

**Fig. 12.** Shear stress $(\sigma)$ is plotted as a function of time in (a) for relaxation during waiting step $(t_w)$ at $Wi = 0$ after pre-shear at $Wi = 20$, and relaxation modulus $(G)$ is plotted as a function of $t - t_w$ obtained after application of step strain, $\gamma = 1$, at different values of $t_w = 10\text{-}70$ (b) and $t_w = 70\text{-}1000$ (c) using viscoelastic aging model by Joshi [50]. The arrows on the figure (b) and (c)

indicates the direction of evolution of relaxation modulus $(G)$ with increasing value of $t_w$ as a function of $t-t_w$.

We next perform identical experiments on the 1.5 wt.% PEO solution and obtain the corresponding theoretical predictions using the FENE-P model. The relaxation of stress in both the systems is shown in Figs. 13 and 14 (a), respectively. As expected, both the 1.5 wt.% PEO solution and the FENE-P model undergo complete relaxation. The application of step-strain over this period shows that lesser stress gets induced in respective systems with increase in $t_w$ as correspondingly shown in the insets of Figs. 13 and 14 (a). However, as $t_w$ becomes comparable to the relaxation time of the solution, the initial modulus becomes independent of $t_w$. Interestingly, unlike that observed for clay dispersions as well as viscoelastic aging model, the relaxation curves do not cross each other. The limiting case of the viscoelastic aging model is the no aging limit of the (time independent) linear Maxwell model. For the flow protocol considered in this section, the Maxwell model can be solved analytically. The stress upon cessation of steady shear flow is given by:

$$\sigma = \exp(-t_w), \tag{8}$$

where $\sigma$ is the dimensionless stress normalized using the steady-state shear stress ($\sigma_0^*$) associated with the pre-shear while $t_w$ is the dimensionless waiting time normalized by the relaxation time of the Maxwell model. We plot the dimensionless stress $\sigma$ as a function of $t_w$ for the Maxwell model in Fig. 14(b). Over the course of relaxation, the Maxwell model is subjected to the step strain $\gamma_1$, at a different value of waiting time $(t_w)$ such that the corresponding stress relaxation modulus is given by:

$$G(t-t_w, t_w) = \left(\sigma\sigma_0^*/G\gamma_1\right) = \left(\sigma_0^*/G\gamma_1\right)\exp(-(t-t_w)-t_w) + \exp(-(t-t_w)), \tag{9}$$

where $G$ is the modulus associated with the Maxwell model. In the inset of Fig. 14(b), we plot $G(t-t_w, t_w)$ as a function of $t-t_w$ for different values of $t_w$. It can be seen that the relaxation modulus curve shifts to lower modulus levels with increase in waiting time, thereby qualitatively corroborating with the behavior of the 1.5 wt.% PEO solution and the FENE-P model. The results of the creep experiments for 1.5 wt.% PEO solution, FENE-P model and linear time-independent Maxwell model are plotted in Fig. S3 and S4 of supplementary information. It can be seen that for $t_w^* < O(\tau^*)$, creep compliance increases with $t_w$ and subsequently it becomes independent of $t_w$ for $t_w^* > O(\tau^*)$. Qualitatively this behavior collaborates with that observed for relaxation modulus shown in Figs. 13 and 14.

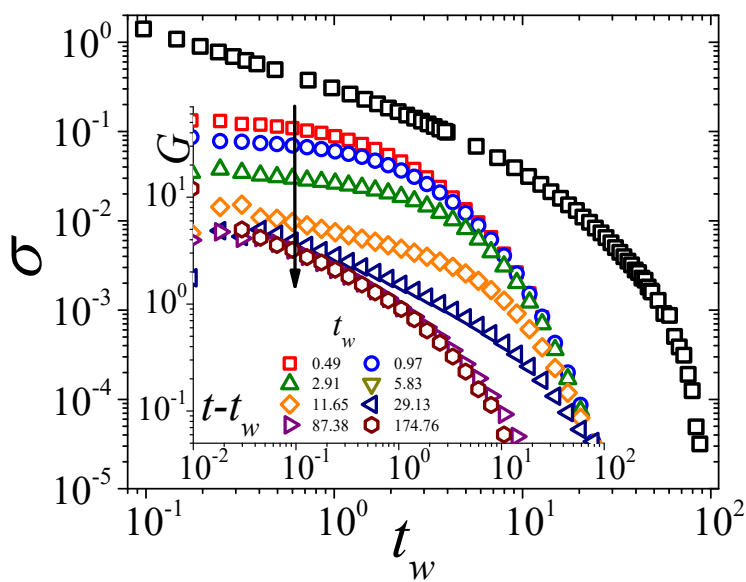

**Fig. 13.** Dimensionless shear stress $(\sigma)$ is plotted against time for relaxation during the waiting step $(t_w)$ at $\dot{\gamma}^* = 0$ after pre-shear at $\dot{\gamma}^* = 10\,\text{s}^{-1}$ for 60 s, and the evolution of dimensionless relaxation modulus $(G)$ is plotted in the inset for a step strain of 1 % for different waiting times $t_w$ for 1.5 wt.% PEO solution. Shear stress and relaxation modulus are non-dimensionalised using $G'$ value at crossover point in the frequency sweep plot of the solution (4.36 Pa), and time units are non-dimensionalised using relaxation time of the solution (10.3 s).

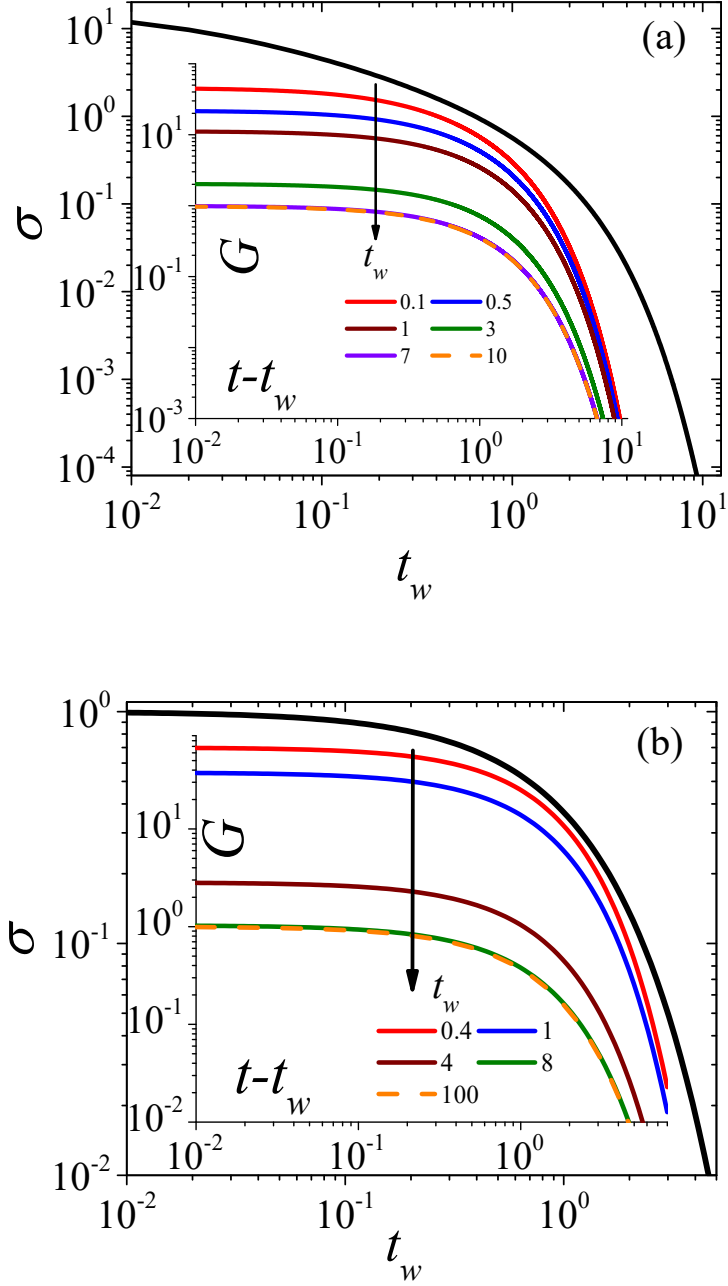

**Fig. 14.** Shear stress ($\sigma$) is plotted against time for relaxation during waiting step ($t_w$) at $Wi = 0$ after pre-shear at $Wi = 100$ for FENE-P model (a) and at $Wi = \sigma_0^*/G$ for the Maxwell model (b), and evolution of relaxation modulus ($G$) is plotted in the inset of (a) for a step strain, $\gamma = 0.1$ for different waiting times $t_w$ for the FENE-P model and in the inset of (b) for a step strain,

$\gamma_1 = 0.01\sigma_0^*/G$ for the Maxwell model. The FENE-P model results are obtained using $b=10^2$ and $\beta=10^{-3}$.

The above discussion clearly suggests that during the timescale over which stress undergoes relaxation following the cessation of steady shear flow, an application of step strain (step stress) induces lower initial stress (higher initial strain) when the material is less relaxed (at lower $t_w$), and the relaxation time of the material itself has not evolved much. This behavior, which is common for physically aging (thixotropic) material and the viscoelastic aging model (over the initial time duration) as well as the viscoelastic material and model, originates from linear viscoelasticity as it is also shown by the Maxwell model. However, for physically aging (thixotropic) materials, interestingly, in the limit of small $t_w$, with increase in $t-t_w$, the relaxation modulus (creep compliance) curves cross over in such a fashion that the curves associated with higher $t_w$ eventually undergo slower relaxation (creep). This behavior can be attributed to increase in relaxation time over the course of relaxation (and creep flow), such that at any $t-t_w$, the higher $t_w$ system has a greater relaxation time. Such behavior is not observed for a viscoelastic material. In a soft glassy material and the viscoelastic aging model, for higher $t_w$, however, the trend seems to be reversing owing to enhancement in relaxation time with waiting time (and also due to increase in elastic modulus with the waiting time in the systems, such as clay dispersion used in this work). In case a soft glassy (thixotropic) material shows complete relaxation of stress (as reported below in Fig. 17 for 2.67 % freshly prepared clay dispersion and Fig. 18 for viscoelastic aging model), application of step strain after attainment of complete relaxation also leads to the waiting time dependence as shown in Figs. 11(c) and 12(c) primarily due to enhancement of relaxation with waiting time. In (non-thixotropic) viscoelastic materials, however, application of step strain (or step stress) beyond complete relaxation does not lead to any waiting time dependence, as relaxation time is constant, and the relaxation modulus (creep compliance) curves at different waiting times fall on top of each other, as shown in Figs. 13 and 14 (Fig. S3 and S4). Consequently, it can be said that if the decay of the relaxation modulus (evolution of creep compliance) with respect to the time elapsed since the application of step strain (step stress) gets slower with increase in waiting time, the material is necessarily a soft glassy or thixotropic material. However, if a system shows a lower value of relaxation modulus (higher value of creep compliance) at any time elapsed since the application of waiting time, it could either a viscoelastic or a physically aging thixotropic system.

We plot a schematic in Fig. 15 that summarizes the above discussion that proposes distinction between thixotropy from (non-thixotropic) viscoelasticity. It should be noted that the step strain experiment is not sufficient to characterize an inelastic thixotropic material as its response will always be identical to viscous material as stress will instantaneously relax on the application of step strain. However, application of step stress would clearly be able to distinguish inelastic thixotropic response from non-thixotropic viscous response. It should be noted that the proposal mentioned below is in accordance with the behavior observed for thixotropic (physically aging) systems such as clay dispersions, microgel paste, hard sphere suspension, and for viscoelastic systems such as PEO solution, FENE-P model and linear Maxwell model. Certainly, more work is needed to establish its unequivocal validity.

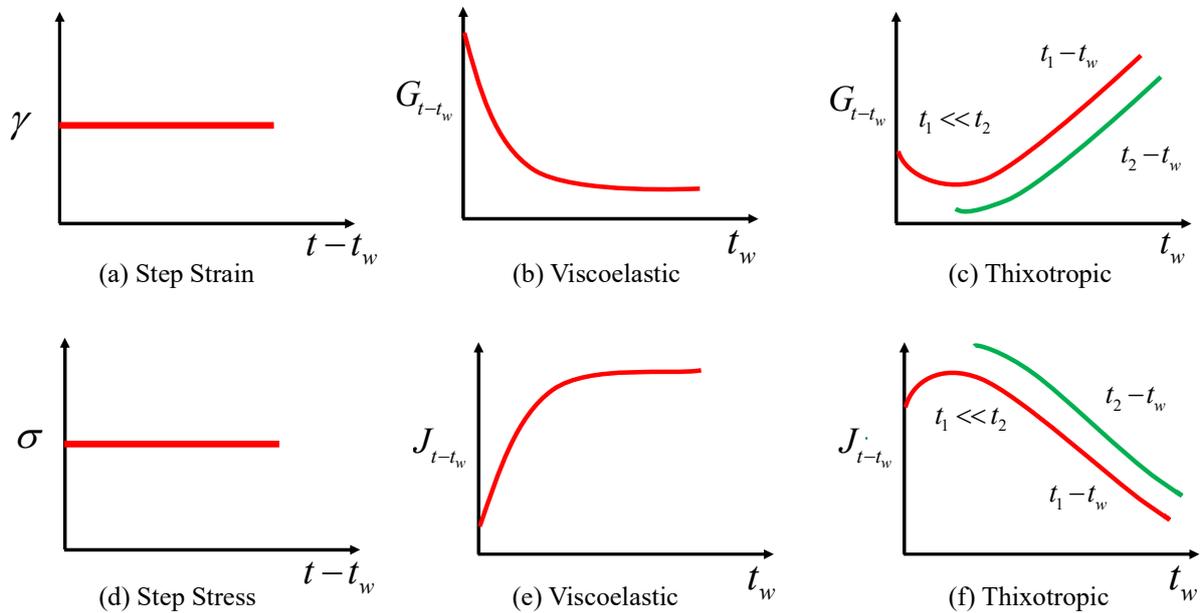

**Fig. 15.** Proposed rheological behavior of a non-thixotropic viscoelastic and thixotropic viscoelastic materials are respectively shown in (b), (e) and (c), (f) for the flow field shown in (a),

(d) applied at different waiting times $t_w$ subsequent to preshear. The parameters $G_{t-t_w}$ and $J_{t-t_w}$ represents relaxation modulus and creep compliance at a fixed value of $t-t_w$ in the stress relaxation modulus and creep compliance evolution and are plotted as a function of $t_w$ (Vertical cut of respective plots on Figs. 11 to 14 and S1 to S4).

### D. Analysis of viscoelastic aging

We now experimentally assess the definition of viscoelastic aging response with respect to other counterparts such as nonthixotropic viscoelastic response and inelastic thixotropic response proposed by Larson and Wei [3]. For this purpose, we use freshly prepared 2.67 wt. % aqueous dispersion of clay having 3 mM NaCl. Subsequent to the mixing and intensely stirring clay with water (for around 30 min), we load the dispersion in the shear cell and subject the same to oscillatory shear flow. At this stage, the viscosity of dispersion is close to water and hence the dispersion does not have any shear/loading history. The corresponding evolution of $G'$ and $G''$ is shown in Fig. 16. In the beginning, the dispersion is in the viscous state as $G'$ is below the detection limit. However, $G''$ increases as a function of time, which is suggestive of increase in viscosity as a function of time (in a purely viscous limit, viscosity $\eta = G''/\omega$). Nevertheless, at a certain point ($t_w^* \approx 4000$ s), $G'$ becomes measurable and undergoes a power law increase. We stop the oscillatory experiment at $t_w^*$=7600 s, and obtain the frequency dependence of dynamic moduli, shown in the inset of Fig. 16. It can be seen that the time at which frequency dependence is obtained, both $G'$ and $G''$ are of the same order of magnitude. Furthermore, the material undergoes time dependent evolution of dynamic moduli (as well as relaxation time). After obtaining the frequency dependence, we subject the same to a constant shear rate ($\dot{\gamma}^*$=0.01 and 0.5 s$^{-1}$) and apply it for $t^* - t_w^*$=720 s. Subsequently, the shear rate is set to $\dot{\gamma}^*$=0 s$^{-1}$. The corresponding stress profile is plotted in Fig. 17. It can be seen that for $\dot{\gamma}^*$=0.01 s$^{-1}$ the stress evolves with time by showing a steady increase, while for $\dot{\gamma}^*$=0.5 s$^{-1}$ a steady-state stress is achieved almost instantaneously. Subsequently, corresponding to $\dot{\gamma}^*$=0 s$^{-1}$, the stress associated with both the shear rates undergoes a complete relaxation (decreases below the sensitivity of the rheometer) without showing any residual stress. The relaxation of stress associated with the system sheared at $\dot{\gamma}^*=0.5$ s$^{-1}$ is expectedly faster as high shear rate breaks the structure formed during the waiting period.

The question is how to categorize the experimental results reported in Fig. 16 and 17. With respect to the distinction proposed by Larson and Wei [3], comparison of the behavior shown in Fig. 17 clearly suggests the behavior to be more close to viscoelastic. However, the previous understanding of this system [58], and the fact that dynamic moduli (as well as relaxation time) grow with time as shown in Fig. 16, suggest that the studied clay dispersion is an eternally aging system that does not reach the equilibrium state over the practically perceivable timescale. The frequency sweep results shown in Fig. 16 clearly suggest the clay dispersion to be viscoelastic with comparable magnitudes of $G'$ and $G''$ over the explored frequency range. Subsequent to start-up shear, the material also undergoes complete relaxation with the timescale of relaxation of the order of $10^3$ s. Therefore, on one hand, the studied clay dispersion appears to be in the viscoelastic aging state; on the other hand, its response is very similar to the non-thixotropic viscoelastic state. Consequently, without a precise experimental way to measure the thixotropic timescale at the instance where shear start-up/stress relaxation study is performed, we are unable to categorize this system as proposed by Larson and Wei [3].

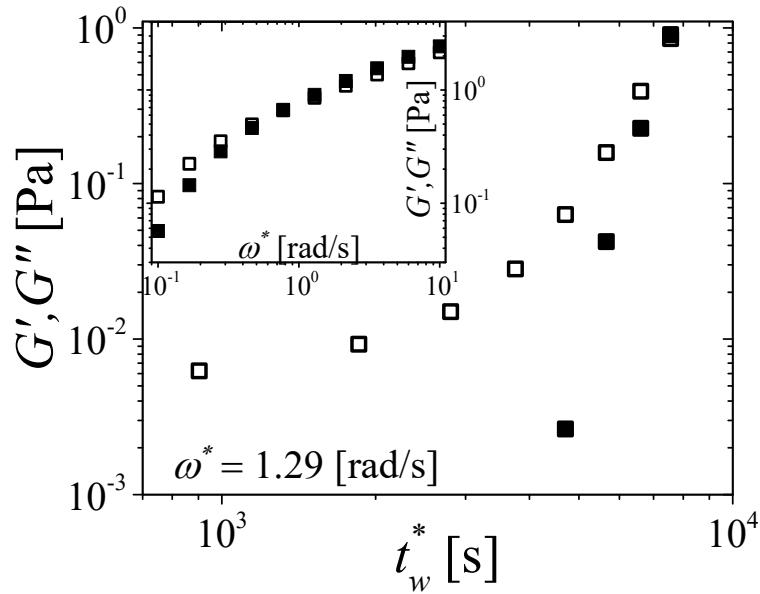

**Fig. 16**. The elastic ($G'$, closed symbols) and viscous ($G''$, open symbols) moduli are plotted as a function of time ($t_w^*$) for a freshly prepared 2.67 wt.% clay with 3 mM NaCl solution at an angular frequency, $\omega^*$=1.29 rad/s. The inset shows the frequency sweep results of the same system obtained at $t_w^*$=7600 s.

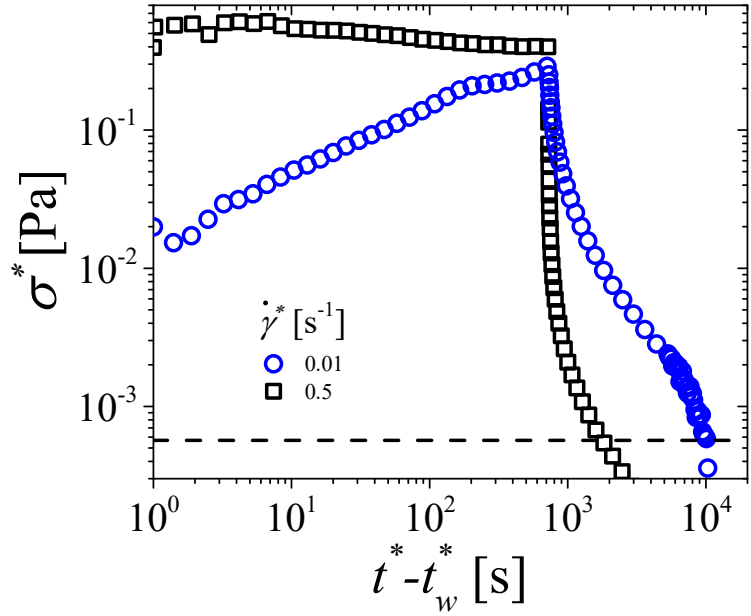

**Fig 17.** Stress response obtained in shear start-up test ($\dot{\gamma}^*$ = 0.01, 0.5 s$^{-1}$ started at $t^* - t_w^*$ = 0) followed by a stress relaxation test ($\dot{\gamma}^*$ = 0 s$^{-1}$) for a freshly prepared 2.67 wt.% clay with 3 mM NaCl solution.

We also solve a viscoelastic aging model [50] with $\mu$ =0.9, 1.0, 1.1 for a case of constant modulus ($B = 0$) for the same protocol wherein starting with $\phi$ =1 (completely rejuvenated state), and after waiting for $t_w = 1$ under quiescent conditions, we subject it to $Wi = 1$ for $t - t_w$=30. Subsequently, we apply $Wi = 0$ and compute the relaxation of stress. The stress response for the step-up shear rate flow and the cessation of shear rate flow are plotted in Fig. 18. We observe that the shear stress increases during the shear start-up and reaches a steady-state. Subsequently, corresponding to $Wi = 0$, the stress undergoes complete relaxation for $\mu$ =0.9 (sub-aging), power law relaxation for $\mu$ =1.0 while incomplete relaxation (approaching residual stress) for $\mu$ =1.1.

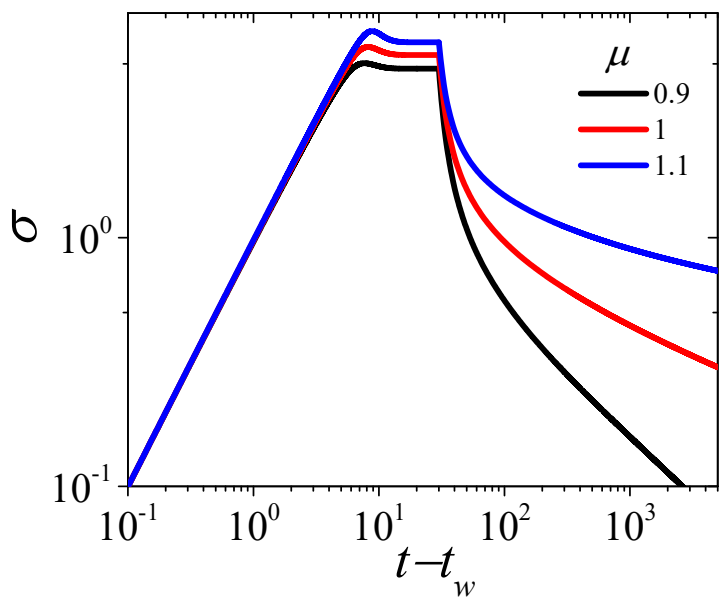

**Fig 18.** Stress response obtained in shear start-up test with $t_w$ =1 and $Wi = 1$ followed by cessation of shear $(Wi = 0)$ leading to relaxation of stress for a viscoelastic aging model with the constant modulus ($B$ =0) for three values of $\mu$ =0.9 (sub aging), 1.0 (full aging), and 1.1 (hyper aging).

Both freshly prepared aqueous clay dispersion and the viscoelastic aging model with $\mu$ =0.9 show a complete relaxation of stress. On the other hand, stress relaxation of 3.5 wt.% clay dispersion as shown in Fig. 11(a) and viscoelastic aging model with $\mu$ =1.1 show incomplete relaxation of stress leading to residual stress in a material. All these systems are not just time evolving systems but also undergo eternal aging. More generally, any time evolving (physically aging) material eventually should undergo complete relaxation of stress if the dependence of relaxation time on waiting time of the same is weaker than unity ($\tau \sim t_w^\mu$ with $\mu < 1$) [50]. This behavior is known as sub-aging behavior [78]. For $\mu = 1$ a material experiences full aging and shows logarithmic relaxation. On the other hand, for $\mu > 1$ a material undergoes hyper-aging and should show a residual stress plateau owing to the inability of the same to completely relax the stress over any time period [50]. Many physically aging soft materials have been observed to undergo sub-aging [46, 79], full aging [42, 43, 80] and hyper-aging dynamics [81]. More specifically, with respect to sub-aging dynamics, Derec and coworkers [46] studied suspensions of 41.7 volume % silica particles having a protective layer of PEO and observed that material undergoes physical aging with $\mu=0.55$. The system undergoes complete relaxation of stress despite it being a viscoelastic (and possibly eternal) aging material. We, therefore, feel that more clarity is needed on distinguishing time dependent viscoelastic materials in various categories as a variety of observations have been reported for such systems. Nonetheless, the present work indicates that thixotropic viscoelastic materials can certainly be distinguished by subjecting the same to the step strain/creep flow fields at different times since cessation of shear melting. After sufficient time has elapsed, the viscoelastic aging/thixotropic materials are observed to show slowing down of the relaxation dynamics as observed for the clay dispersion shown in Fig. 11 and Fig. S1 as well as for various other systems studied in the literature. However, more theoretical and experimental investigation is needed to ascertain whether this behavior is universal.

**V. Summary and concluding remarks**

Owing to the very broad and inclusive definition of thixotropy, often the characteristic features of materials showing non-linear viscoelastic response along with shear thinning can be interpreted as thixotropic, especially during transient evolution. Consequently, it is imperative to have a definitive and unambiguous demarcation between thixotropy and viscoelasticity; this aspect has been debated for the past several decades in the literature. In the absence of any straightforward answer to this question, it is customary to exclude equilibrium viscoelastic materials such as polymeric liquids (solutions and melts) and corresponding constitutive equations, that are universally accepted as non-thixotropic, from the classification of thixotropic materials [2, 3, 29]. In order to determine whether an *a priori* unknown material is thixotropic or viscoelastic, some criteria have been proposed in the literature. The primary objective of the present work is to assess these criteria using universally accepted non-thixotropic viscoelastic systems. To this end, we employ an aqueous 1.5 wt.% PEO solution and the FENE-P model to theoretically predict the behavior. We also investigate a model soft glassy material, an aqueous dispersion of clay, that is known to undergo physical aging and rejuvenation, along with a viscoelastic aging model. While there is some debate on whether to consider soft glassy materials thixotropic [3, 29], clay suspensions have been considered to be thixotropic in a vast majority of the literature [52-57].

We subject the aforementioned systems to various flow protocols proposed in the literature, such as a step-down jump in shear rate, shear start-up at different waiting times, stress relaxation and creep at different waiting times. We observe that in a step-down shear rate protocol, both the aqueous 1.5 wt.% PEO solution and the FENE-P model show stress undershoot that is proposed in the literature as a signature of viscoelastic thixotropic response. Our analysis suggests that this behavior originates from non-linear viscoelasticity coupled with shear thinning response. The shear start-up experiments also show an overshoot whose intensity goes on increasing with increase in waiting times elapsed since pre-shear, for waiting times up to relaxation time of a material. Therefore, if the relaxation time of a material is extremely large (for non-thixotropic viscoelastic materials, the relaxation time may vary from nanoseconds to centuries [3]), the shear start-up experiment at different waiting times cannot distinguish the thixotropic response from a viscoelastic response.

Interestingly, application of step strain and step stress at different waiting times since pre-shear leads to a noteworthy distinction between soft glassy thixotropic systems (aqueous clay dispersion and viscoelastic aging model) in comparison with the non-thixotropic viscoelastic systems (aqueous 1.5 wt.% PEO solution and FENE-P model). We observe that in the latter, the stress relaxation modulus always shifts to lower values of modulus for step strain applied at higher waiting times, for waiting time below the relaxation time of a material. For waiting times above the relaxation time, a viscoelastic material reaches equilibrium and hence relaxation modulus curves superpose on each other. Equivalently, the creep compliance of a viscoelastic material always shifts to a higher value for step stress experiments, if the waiting time is lesser than the relaxation time of a material else it becomes independent of waiting time. For soft glassy thixotropic materials, at small waiting times, the relaxation modulus curves do show faster relaxation at early times with increase in waiting time, but eventually, relaxation gets slower as time increases. At higher waiting times, the overall relaxation gets slower with waiting time and relaxation modulus curves shift to higher modulus values. This latter distinct feature of slowing down of the relaxation dynamics is not shown by the non-thixotropic viscoelastic materials. We regard that this feature that originates from an enhancement in relaxation time with increase in waiting time may be used as a defining criterion that necessarily suggests the presence of thixotropy if the material also undergoes shear rejuvenation. Similarly, creep compliance initially shifts to a higher value at lower waiting time but shifts to a lower value at higher waiting time due to increase in relaxation time. To this end, it would be helpful to study other experimental systems as well as theoretical models to assess their behavior from a thixotropic or viscoelastic point of view. We also study the distinction between viscoelastic systems and viscoelastic aging systems as proposed by Larson and coworkers [3]. We report that, physically aging viscoelastic systems (more specifically, eternally aging systems), depending upon whether the evolution of relaxation time is weaker or stronger than waiting time, show a variety of rheological responses. Consequently, it is difficult to distinguish between viscoelastic behavior and viscoelastic aging behavior. However, the response to the step strain (or step stress) as a function of waiting time can still be used as a distinguishing factor between non-thixotropic viscoelastic and viscoelastic aging.

In order to adopt the above-mentioned criterion as a differentiating measure, the question still remains whether soft glassy materials that undergo physical aging (viscoelastic aging systems wherein relaxation time increases as a function of waiting time) and show shear rejuvenation (decrease in relaxation time and/or viscosity with time under application of flow) can be considered as thixotropic or not. While the rheological behavior of physically aging soft glassy materials unequivocally satisfies the definition of thixotropy, and has been considered as thixotropic by predominant fraction of researchers in this field, a recent proposal by Larson and coworkers [3] deviates from this viewpoint. This is because for soft glassy materials, the relaxation time is usually of the order of thixotropic timescale over most of the ages (if not all the ages), that according to the definition proposed by Larson and coworkers, excludes soft glassy materials at advanced age from thixotropic classification. Nonetheless, the motivation of this article is not to question the proposal by Larson and coworkers, though we do consider that the soft glassy materials are indeed thixotropic in nature. We also do not attempt to redefine thixotropy so that it unambiguously differentiates itself from the traditional definition of viscoelasticity. The purpose of the present work is to assess various criteria proposed in the literature for commonly known non-thixotropic viscoelastic materials – polymeric liquids. We observe that while the criteria proposed earlier are not successful in distinguishing the two, the step-strain and the step stress protocol at different waiting times elapsed since the cessation of the deformation field, do lead to definitive demarcation. By way of further scrutiny, more work is required to make this criterion more robust and universal.

**Supplementary Online Material:** See Supplementary Material for creep results at different waiting times for all the studied systems.

**Acknowledgment**: We acknowledge financial support from the Science and Engineering Research Board, Government of India. We also thank Prof. Jan Mewis and Prof. Francisco Rubio-Hernandez for their insightful comments. We are particularly grateful to Prof. Ronald Larson for discussion on several aspects of thixotropy that has helped us in clarifying various critical issues.

# Supplementary Information:

# Distinguishing Thixotropy from Viscoelasticity

Mayank Agarwal,[a] Shweta Sharma,[a] V. Shankar*, Yogesh M. Joshi*

Department of Chemical Engineering,

Indian Institute of Technology Kanpur, Kanpur 208016. INDIA.

* Corresponding authors, email id: vshankar@iitk.ac.in and joshi@iitk.ac.in

[a] Both authors contributed equally to this manuscript.

After pre-shearing the samples, we also perform the creep experiments at different waiting times. We perform these experiments on 3.5 wt.% clay dispersion and 1.5 wt.% PEO solution. We also solve the viscoelastic aging model, linear Maxwell model, and FENE-P model that have already been discussed in the main text. In a typical protocol we pre-shear the experimental systems/theoretical models as mentioned in the main text. After stopping pre-shear, we subject the same to $\dot{\gamma}^*=0$ for different waiting times. Subsequently we apply step stress and monitor evolution of compliance.

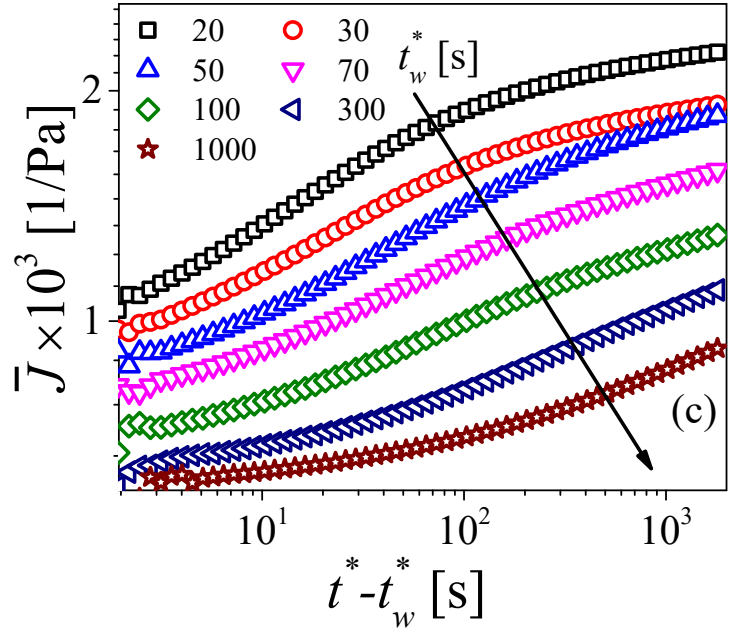

**Fig. S1.** (a) Shear stress relaxation subsequent to the cessation of steady shear flow at (50 s$^{-1}$ applied for 300 s) for 3.5 wt.% clay dispersion. The stress undergoes relaxation and reaches a plateau for $t_w^* > 10$ s. The sample is subjected to a step stress of 10 Pa at different waiting times after cessation of steady shear flow. The corresponding evolution of creep compliance is plotted for different waiting times $\left(t_w^*\right)$ = 0.105 to 0.4 (b), 0.3 to 20s (the inset of Fig. (b)) and = 20 to 1000s (c). The arrows on the figure (b) and (c) indicates the direction of increasing value of $t_w^*$. The inset of figure (b) shows that the evolution of $\bar{J}$ with $t^* - t_w^*$ for different $t_w^*$ cross each other because of increase in relaxation time in the system, and the crossover time decreases with increase in $t_w^*$ as indicated by the highlighted data points. On the other hand, at higher $t_w^*$, the $\bar{J}$ at any $t^* - t_w^*$ decrease with $t_w^*$ as shown in figure (c).

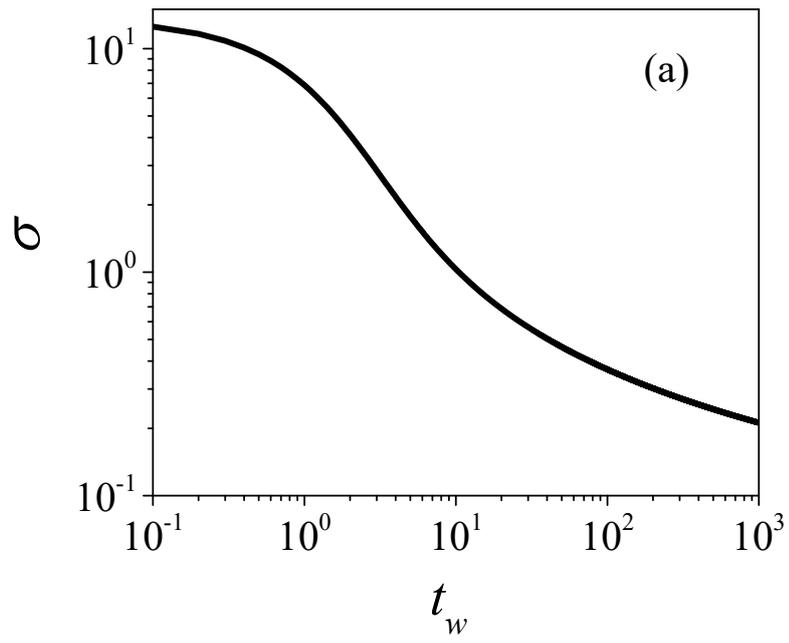

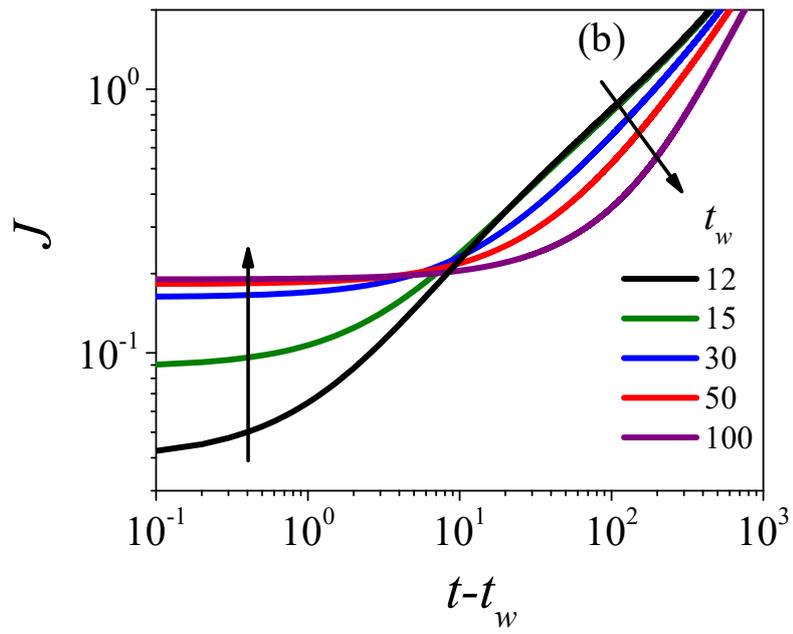

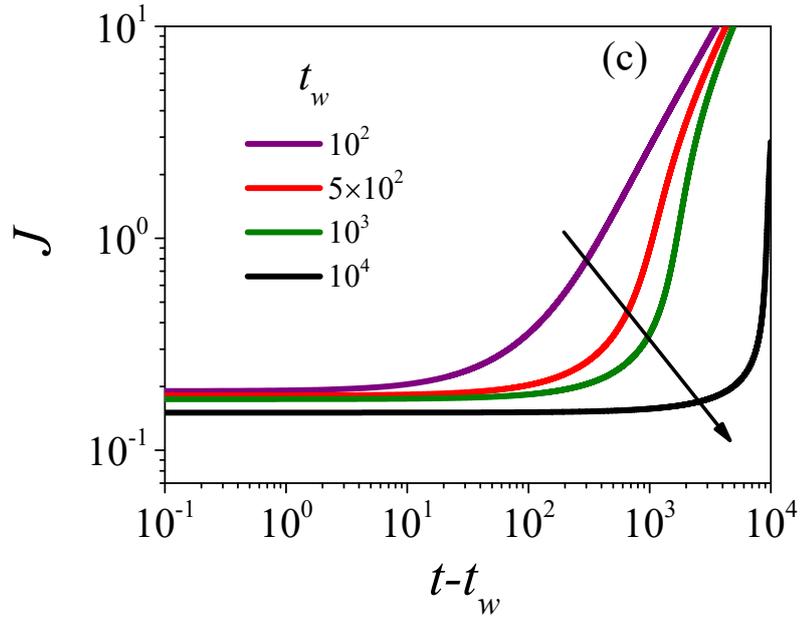

**Fig. S2.** Shear stress $(\sigma)$ is plotted as a function of time for relaxation during the waiting step at $Wi = 0$ subsequent to pre-shear at $Wi = 10$ (Fig (a)) for the viscoelastic aging model. Creep compliance $(J)$ obtained after application of step stress $(\sigma = 1)$ is plotted as a function of $t - t_w$, at different waiting times $t_w = 12 - 10^2$ (Fig (b)) and $t_w = 10^2 - 10^4$ (Fig (c)). These results are obtained using model parameters: $A = 2.5$, $B = \ln 10$ and $\mu = 1.01$. The arrows on the figure (b) and (c) indicates the direction of increasing value of $t_w$. It can be seen that at higher $t_w$, the $J$ at any $t - t_w$ decrease with $t_w$ as shown in figure (c).

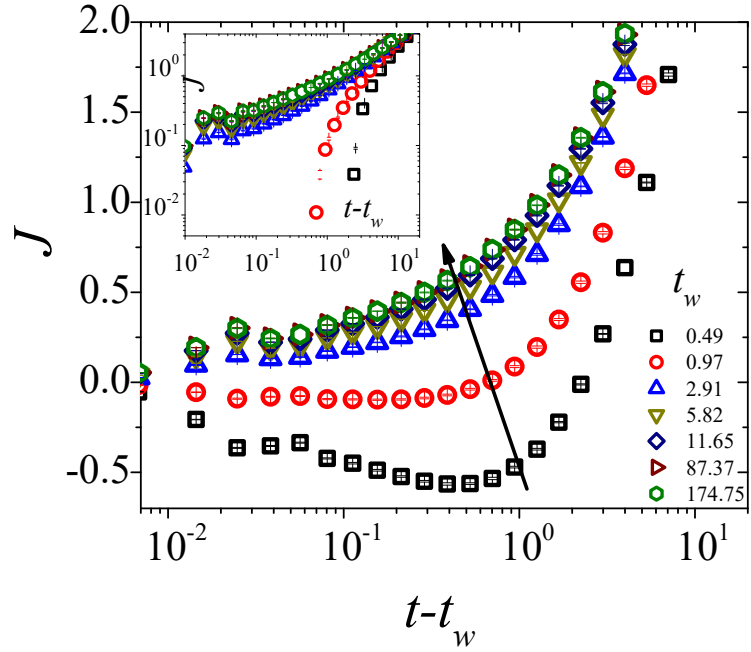

**Fig S3.** The dimensionless creep compliance $(J)$ of 1.5 wt.% PEO solution is plotted as a function of dimensionless time $(t - t_w)$ for an applied step stress of 1 Pa after different dimensionless waiting times $(t_w)$, as shown in the legend. The inset shows the same data on a log-log scale plot. The arrow shows the direction of increasing value of waiting time. The creep compliance is observed to be negative due to elastic effects at the lower value of $t_w$, if the applied stress is lesser than the initial value of stress during creep experiment. The stress relaxation during waiting step can be referred from Fig. 13. The stress units are non-dimensionalised using $G'$ value at crossover point (= 4.36 Pa) in the dynamic moduli - frequency plot of the solution shown in Fig.3(b). The time units are non-dimensionalised using relaxation time of the solution (10.3 s).

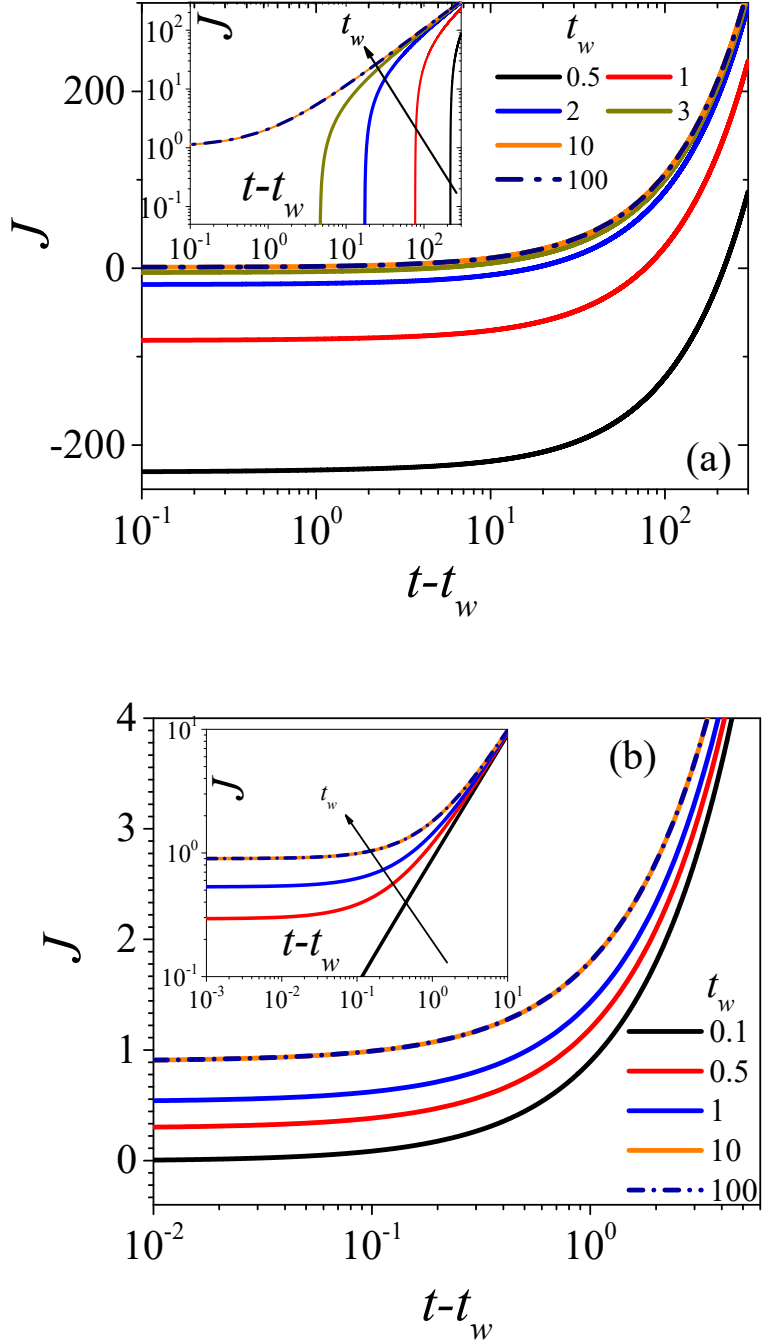

**Fig. S4.** Creep compliance $(J)$ is plotted as a function of $t-t_w$, wherein for FENE-P model the pre-shear was carried out at $Wi=100$ (a), while the Maxwell model was subjected to pre-shear at $Wi=\sigma_0^*/G$ (b). During the waiting period ($Wi=0$) both the models undergo stress relaxation. The dimensionless applied stress is 0.01 for the FENE-P model and $0.9\sigma_0^*/G$ for the Maxwell model. The inset of figure (a) and (b) shows the same results on a logarithmic scale. The arrow shows the direction of increasing value of waiting time. The FENE-P model

results are obtained for: $b = 10^2, \beta = 10^{-3}$. The creep compliance is observed to be negative due to elastic effects at the lower value of $t_w$, if the applied stress is lesser than the initial value of stress during creep experiment.